\documentclass[11pt, reqno]{amsart}
\usepackage{amsfonts}
\usepackage{amssymb}
\usepackage{latexsym}
\usepackage[final]{hyperref}
\usepackage[]{enumerate}
\usepackage{verbatim}
\usepackage[english]{babel} 
\usepackage{blindtext}
\usepackage[nohyperlinks,nolist]{acronym}
\usepackage{listings}
\usepackage{multirow}
\usepackage{dsfont}

\usepackage{mathtools}

\usepackage[table]{xcolor}%
\newcommand{\major}{\cellcolor{blue!100}}
\newcommand{\minor}{\cellcolor{blue!25}}

\usepackage{graphicx}       
\usepackage[utf8]{inputenc} 
\usepackage{amsmath}

\usepackage{tikz}
\usepackage{cmap}
\usepackage{ulem}
\usepackage{soul}
\usepackage{comment}
\usepackage{fancybox}
\usepackage{relsize}
\usepackage{pifont}
\usepackage{subcaption}

\usetikzlibrary{arrows.meta, decorations.markings}
\usepackage{pgfplots}
\pgfplotsset{compat=1.18}
\usetikzlibrary{arrows.meta}

\renewcommand{\theequation}{\arabic{section}.\arabic{equation}}

\newcommand{\Abs}[1]{\left\vert #1 \right\vert}

\newcommand{\norm}[1]{\left\Vert #1 \right\Vert}

\DeclareMathOperator{\trace}{tr}

\DeclareMathOperator{\Span}{Span}

\DeclareMathOperator{\Rank}{Rank}
\DeclareMathOperator{\Image}{Im}
\DeclareMathOperator{\Ker}{Ker}

\newtheorem{theorem}{Theorem}
\newtheorem{proposition}{Proposition}

\theoremstyle{definition}
\newtheorem{definition}{Definition}

\theoremstyle{remark}
\newtheorem{remark}{Remark}

\begin{acronym}
\acro{AO}{atomic orbital}
\acro{API}{Application Programmer Interface}
\acro{AUS}{Advanced User Support}
\acro{BEM}{Boundary Element Method}
\acro{BO}{Born-Oppenheimer}  
\acro{CBS}{complete basis set}
\acro{CC}{Coupled Cluster}
\acro{CTCC}{Centre for Theoretical and Computational Chemistry}
\acro{CoE}{Centre of Excellence}
\acro{DC}{dielectric continuum}  
\acro{DCHF}{Dirac-Coulomb Hartree-Fock}  
\acro{DFT}{density functional theory}  
\acro{DIIS}{direct inversion in the iterative subspace}  
\acro{DKH}{Douglas-Kroll-Hess}
\acro{EFP}{effective fragment potential}
\acro{ECP}{effective core potential}
\acro{EU}{European Union}
\acro{FFT}{Fast Fourier Transform}
\acro{GGA}{generalized gradient approximation}
\acro{GPE}{Generalized Poisson Equation}
\acro{GTO}{Gaussian Type Orbital}
\acro{HF}{Hartree-Fock}  
\acro{HPC}{high-performance computing}
\acro{Hylleraas}[HC]{Hylleraas Centre for Quantum Molecular Sciences}
\acro{IEF}{Integral Equation Formalism}
\acro{IGLO}{individual gauge for localized orbitals}
\acro{KB}{kinetic balance}
\acro{KS}{Kohn-Sham}
\acro{LAO}{London atomic orbital}
\acro{LAPW}{linearized augmented plane wave}
\acro{LDA}{local density approximation}
\acro{MAD}{mean absolute deviation}
\acro{maxAD}{maximum absolute deviation}
\acro{MM}{molecular mechanics}  
\acro{MCSCF}{multiconfiguration self consistent field}
\acro{MPA}{multiphoton absorption}
\acro{MRA}{multiresolution analysis}
\acro{MSDD}{Minnesota Solvent Descriptor Database}
\acro{MW}{multiwavelet}
\acro{NAO}{numerical atomic orbital}
\acro{NeIC}{nordic e-infrastructure collaboration}
\acro{KAIN}{Krylov-accelerated inexact Newton}
\acro{NMR}{nuclear magnetic resonance}
\acro{NP}{nanoparticle}  
\acro{NS}{non-standard}  
\acro{OLED}{organic light emitting diode}
\acro{PAW}{projector augmented wave}
\acro{PBC}{Periodic Boundary Condition}
\acro{PCM}{polarizable continuum model}
\acro{PW}{plane wave}
\acro{QC}{quantum chemistry}  
\acro{QM/MM}{quantum mechanics/molecular mechanics}  
\acro{QM}{quantum mechanics}  
\acro{RCN}{Research Council of Norway}
\acro{RMSD}{root mean square deviation}
\acro{RKB}{restricted kinetic balance}
\acro{SC}{semiconductor}
\acro{SCF}{self-consistent field}
\acro{STSM}{short-term scientific mission}
\acro{SAPT}{symmetry-adapted perturbation theory}
\acro{SERS}{surface-enhanced raman scattering}
\acro{WPREL}[WP1]{Work Package 1}
\acro{WPROP}[WP2]{Work Package 2}
\acro{WPAPP}[WP3]{Work Package 3}
\acro{WP}{Work Package}  
\acro{X2C}{exact two-component}
\acro{ZORA}{zero-order relativistic approximation}
\acro{ae}{almost everywhere}
\acro{BVP}{boundary value problem}
\acro{PDE}{partial differential equation}
\acro{RDM}{1-body reduced density matrix}
\acro{SCRF}{self-consistent reaction field}
\acro{IEFPCM}{Integral Equation Formalism \ac{PCM}}
\acro{FMM}{fast multipole method}
\acro{DD}{domain decomposition}
\acro{TRS}{time-reversal symmetry}
\acro{SI}{Supporting Information}
\acro{DHF}{Dirac--Hartree--Fock}
\acro{MO}{molecular orbital}
\end{acronym}

\title
[
    Multi-configurational optimization
]
{
    Newton optimization for the Multiconfiguration Self Consistent Field method at the basis set limit: closed-shell two-electron systems.
}

\author[Dinvay]{Evgueni Dinvay}
\author[Vikhamar-Sandberg]{Rasmus Vikhamar-Sandberg}

\email{ evgueni.dinvay@gmail.com }

\address
{
    Department of Chemistry
    \\
    UiT The Arctic University of Norway
    \\
    PO Box 6050 Langnes
    \\
    N-9037 Tromsø
    \\
    Norway
}

\date{\today}

\begin{document}

\begin{abstract}
The multiconfiguration self-consistent field (MCSCF) method is revisited with a specific focus on two-electron systems for simplicity.
The wave function is represented as a linear combination of Slater determinants.
Both the orbitals and the coefficients of this configuration interaction expansion are optimized according to the variational principle within the Lagrangian formalism, using a Newton optimization scheme.
This reduces the MCSCF problem to solving a particular differential Newton system, which can be discretized with multiwavelets and solved iteratively.
\end{abstract}

\maketitle

\section{Introduction}
\setcounter{equation}{0}

We present the first implementation of a Newton optimization scheme for the \ac{MCSCF} method within the framework of \ac{MRA} using \acp{MW}. 
Multiconfigurational approaches are essential for capturing correlation effects in quantum chemistry. However, the optimization of \ac{MCSCF} wavefunctions remains a challenging nonlinear problem due to the coupled dependence on both configurational coefficients and orbitals. It is widely recognized that second-order methods, in particular Newton-type algorithms, provide the most efficient route to convergence when reliable second-derivative information is available.

In conventional formulations, the \ac{MCSCF} energy is expressed in terms of a finite basis representation, and the Newton step is derived from explicit second derivatives with respect to both configuration interaction coefficients and orbital rotation parameters \cite{Helgaker_Jorgensen_Olsen}.
In contrast, within a multiwavelet framework, one operates in a virtually infinite basis set, where such parametrizations are neither natural nor numerically advantageous.
This necessitates a reformulation of the optimization problem
at the functional (basis independent) level.

In this work, we adopt a Lagrangian formulation of the \ac{MCSCF} problem and derive the corresponding Newton equations directly in function space.
This approach enables the incorporation of Green's function techniques and leads naturally to an integral formulation of the Newton step.
The use of resolvent operators plays a central role in this construction and is closely aligned with the multiresolution representation.

For this prototype study we restrict ourselves to the case of a two-electron closed-shell system.
While this setting allows for a transparent exposition of the method, the formulation can be generalized and extended to systems with an arbitrary number of electrons and general spin symmetry.

We emphasize that the present formulation is closely related to recent developments in geometric optimization.
In particular, the variational parameters of the \ac{MCSCF} problem naturally
reside on nonlinear manifolds due to orthonormality constraints.
While the present work is formulated in a Lagrangian framework,
it can be viewed as a precursor to a fully Riemannian treatment of the problem.
A consistent definition of the energy gradient in the Riemannian sense has only recently been introduced
for single-determinant methods \cite{Dinvay2026gradient_descent_hartree_fock},
and its extension to the multiconfigurational setting is an important direction for future work.
In the present work, however,
we employ tools that are more familiar to specialists in quantum chemistry,
namely, Lagrange's multiplier theorem for the constrained optimization
and the Hilbert space of square integrable functions.

Our approach allows one to perform \ac{MCSCF} calculations
of both ground and excited states within the framework of \ac{MRA}.
Here, we note that this adaptive discretization technique is particularly useful in computational chemistry, where it effectively handles the singular Coulomb interaction between an electron and the nucleus, as well as the resulting cusp in orbitals
\cite{
    Frediani_Fossgaard_Fla_Ruud,
    Harrison_Fann_Yanai_Gan_Beylkin2004,
    Jensen_Saha_elephant2017,
    Yanai_Fann_Gan_Harrison_Beylkin2004}.
Moreover, its success in chemistry applications is largely due to the approximation of the resolvent:
\begin{equation}
\label{resolvent_approximation}
    H(\mu)
    =
    \left( - \Delta + \mu^2 \right)^{-1}
    \approx
    \sum_{j} c_j e^{t_j \Delta}
\end{equation}
based on the heat semigroup \cite{Harrison_Fann_Yanai_Beylkin2003}.
Multiresolution quantum chemistry is rapidly evolving, and we refer readers to a comprehensive review for a broader perspective \cite{Bischoff2019}.

We employ the Lagrange's multiplier theorem for constrained minimization of the \ac{MCSCF} energy.
Although this approach is less common in quantum chemistry
--
since stationary points of the Lagrangian $\mathcal L$ are saddle points
--
it provides a natural framework for incorporating Green's operators.
As a result, the differential Newton system can be reformulated in an integral form,
which is particularly well suited for implementation within the multiresolution framework.
An alternative \ac{MRA} approach to \ac{MCSCF} problem
can be seen in recent works
\cite{Nibbi_Frediani_Dinvay_Mendl2025, Valeev_Harrison_Holmes_Peterson_Penchoff2023}.

\section{Notations}
\setcounter{equation}{0}

All the formulas are given in atomic units.
Throughout the text we use the notation $\varphi, \varphi_i, \delta \varphi_i$ for functions in $L^2 \left( \mathbb R^3 \right)$, which are referred to as orbitals.
We focus on closed-shell two-electron systems. A Slater determinant associated with a spatial orbital $\varphi_m$ is denoted by
\(
    \left| m \overline{m} \right \rangle,
\)
indicating spin-up and spin-down electrons occupying the same spatial orbital.
The one-electron Hamiltonian is defined as
\[
    h
    =
    - \frac 12 \Delta + V_{\text{nuc}}.
\]

We employ the standard notation
\begin{equation*}
    (i|h|j)
    =
    \int
    \varphi_i(x) h \varphi_j(x)
    \, dx
\end{equation*}
for one-electron integrals and
\begin{equation*}
    (ij|kl)
    =
    \int
    \frac{1}{|x - y|}
    \varphi_i(x) \varphi_j(x)
    \varphi_k(y) \varphi_l(y)
    \, dx dy
\end{equation*}
for two-electron Coulomb integrals \cite{Szabo_Ostlund}.
We frequently encounter convolution expressions of the form
\[
    J (i \, \delta k)
    =
    \frac{1}{|x|} * (\varphi_i \, \delta \varphi_k)
    ,
\]
which is a function in $L^{3*} \left( \mathbb R^3 \right)$,
known as weak $L^3$-space,
by the Hardy--Littlewood--Sobolev lemma.
In the computational chemistry literature,
such expressions are naturally associated with Coulomb and exchange potentials,
which motivates this notation.

Finally, we introduce the resolvent operator
\begin{equation}
\label{resolvent}
    R_i
    =
    \left(
        - \frac {c_i^2}{2} \Delta - \varepsilon_{ii}
    \right)^{-1}
    =
    \frac{2}{c_i^2}
    H\!\left(
        \sqrt{ - \frac {2 \varepsilon_{ii}}{c_i^2} }
    \right)
    ,
\end{equation}
where the orbital energies $\varepsilon_{ii}$ are assumed to be negative.
This ensures that the orbital energies lie in the resolvent set of the kinetic energy.

In practical computations, this condition may be temporarily violated during early iterations.
Such a situation indicates that the current iterate is far from convergence.
In this case, one may enforce negativity by replacing $\varepsilon_{ii}$ with $-\varepsilon_{ii}$.
As will become clear below, this procedure is consistent with the standard level-shift strategy
used to stabilize the convergence in Newton-type methods.

In general, an optimization problem can be recast
as a stationary point equation
\begin{equation}
\label{general_stationary_equation}
    \mathcal R(w) = 0
\end{equation}
with some $\mathcal R : W \to W$,
where the space $W$ may contain functions, coefficients and Lagrange multipliers.
Newton's method can be described as follows.
For a given iterate $w = w_n$ we solve the Newton's equation
\begin{equation}
\label{general_newton}
    d \mathcal R(w) \delta w
    =
    - \mathcal R(w)
\end{equation}
with respect $\delta w$.
This is a linear equation that we solve iteratively.
Note that in a basis type method one could potentially invert the Jacobian $d \mathcal R(w)$,
though for this the basis size should be prohibitively small.
Then the update $\delta w$ gives the new iterate $w_{n + 1} = w_n + \delta w$,
which in turn defines the new Newton's equation \eqref{general_newton},
now with $w = w_{n + 1}$ and new unknown $\delta w$.
In other words the complete iteration procedure consists of two loops.
The innermost loop is solved with the help of fixed-point iterations.
For this purpose we rewrite \eqref{general_newton} in
the so called self-consistent form
\begin{equation}
\label{general_self_consistent_form}
    \delta \varphi
    =
    \mathbb F( \delta \varphi; w )
    ,
\end{equation}
where we separate the orbital update $\delta \varphi$
from the full update vector $\delta w$.
In \eqref{general_self_consistent_form}
the point $w$ is a fixed parameter,
while $\delta \phi$ is unknown.
One iterates
\(
    \delta \varphi_{n + 1}
    =
    \mathbb F( \delta \varphi_n; w )
\)
until self-consistency.
It can be accelerated with \ac{DIIS},
an extrapolation technique also known as Pulay mixing
\cite{Pulay1980, Rohwedder_Schneider2011}.
Concrete examples of $\mathcal R$ in \eqref{general_newton} and
$\mathbb F$ in \eqref{general_self_consistent_form}
are provided below in the text.

\section{Two configurations}
\setcounter{equation}{0}

Before considering the general multideterminant problem,
we would like to focus on the simplest possible case of
linear combination of two determinants $\left| 1 \overline{1} \right \rangle$
and $\left| 2 \overline{2} \right \rangle$,
as it is formulated in \cite{Szabo_Ostlund}.
The corresponding CI coefficients satisfy
\[
    c_1^2 + c_2^2 = 1
\]
and the spatial orbitals $\varphi_1, \varphi_2$ are orthonormalized.
The amount of variables can be reduced by introducing an angle $\theta$,
so that $c_1 = \cos \theta$ and $c_2 = \sin \theta$.
As the Hartree-Fock theory gives the dominating configuration,
one anticipates the angle $\theta$ to be small.
The total energy has the form
\[
    E(\varphi_1, \varphi_2, \theta)
    =
    \cos^2 \theta
    ( 2(1 |h| 1) + (11|11) )
    +
    \sin^2 \theta
    ( 2(2 |h| 2) + (22|22) )
    +
    \sin 2\theta
    (12|12)
    .
\]
Introduce the Lagrangian
\[
    \mathcal L
    =
    E - 2 \sum_{i, j  = 1}^2 \varepsilon_{ij} \left( \int \varphi_i \varphi_j - \delta_{ij} \right)
\]
with symmetric matrix $\varepsilon$.
It is necessary to impose its symmetry,
because we have only 3 orthogonality restrictions on the orbitals.
An \ac{MCSCF} wave function corresponds to the global minimum of the energy,
and so a stationary point of $\mathcal L$,
namely, a solution of the equation $\nabla \mathcal L = 0$.
The gradient consists of the variational derivatives
\begin{equation}
\begin{aligned}
    \frac 14 \frac{\delta \mathcal L}{\delta \varphi_1}
    =
    c_1^2
    \left(
        h + \frac 1{|x|} * \varphi_1^2
    \right)
    \varphi_1
    +
    c_1 c_2
    \left(
        \frac 1{|x|} * (\varphi_1 \varphi_2)
    \right)
    \varphi_2
    -
    \varepsilon_{11} \varphi_1
    -
    \varepsilon_{12} \varphi_2
    \\
    \frac 14 \frac{\delta \mathcal L}{\delta \varphi_2}
    =
    c_2^2
    \left(
        h + \frac 1{|x|} * \varphi_2^2
    \right)
    \varphi_2
    +
    c_1 c_2
    \left(
        \frac 1{|x|} * (\varphi_1 \varphi_2)
    \right)
    \varphi_1
    -
    \varepsilon_{12} \varphi_1
    -
    \varepsilon_{22} \varphi_2
\end{aligned}
\end{equation}
and the partial derivatives with respect to
\(
    \theta, \varepsilon_{11}, \varepsilon_{22}, \varepsilon_{12}
    .
\)
In particular,
$\theta$-derivative:
\[
    \frac{\partial \mathcal L}{\partial \theta} (\varphi_1, \varphi_2, \theta)
    =
    \frac{\partial E}{\partial \theta} (\varphi_1, \varphi_2, \theta)
\]
where
\[
    \frac{\partial E}{\partial \theta} (\varphi_1, \varphi_2, \theta)
    =
    - \sin 2 \theta
    ( 2(1 |h| 1) + (11|11) )
    +
    \sin 2 \theta
    ( 2(2 |h| 2) + (22|22) )
    +
    2 \cos 2\theta
    (12|12)
    .
\]
This brings us to the following system of integro-differential equations
\begin{equation}
\begin{aligned}
    c_1^2
    \left(
        h + \frac 1{|x|} * \varphi_1^2
    \right)
    \varphi_1
    +
    c_1 c_2
    \left(
        \frac 1{|x|} * (\varphi_1 \varphi_2)
    \right)
    \varphi_2
    =
    \varepsilon_{11} \varphi_1
    +
    \varepsilon_{12} \varphi_2
    \\
    c_2^2
    \left(
        h + \frac 1{|x|} * \varphi_2^2
    \right)
    \varphi_2
    +
    c_1 c_2
    \left(
        \frac 1{|x|} * (\varphi_1 \varphi_2)
    \right)
    \varphi_1
    =
    \varepsilon_{12} \varphi_1
    +
    \varepsilon_{22} \varphi_2
\end{aligned}
\end{equation}
complemented by
\[
    - \sin 2 \theta
    ( 2(1 |h| 1) + (11|11) )
    +
    \sin 2 \theta
    ( 2(2 |h| 2) + (22|22) )
    +
    2 \cos 2\theta
    (12|12)
    =
    0
\]
and by the orthonormality condition.
It is very difficult problem demanding advance numerical techniques to be exploited.
It is commonly known \cite{Helgaker_Jorgensen_Olsen}
that the Newton's optimization is the most suitable method here. 
We introduce
\[
    \mathcal R( \varphi_1, \varphi_2, \theta, \varepsilon_{11}, \varepsilon_{22}, \varepsilon_{12} )
    =
    \left(
        \frac 14 \frac{\delta \mathcal L}{\delta \varphi_1}
        ,
        \frac 14 \frac{\delta \mathcal L}{\delta \varphi_2}
        ,
        \frac{\partial \mathcal L}{\partial \theta}
        ,
        \int \varphi_1 \varphi_1 - 1
        ,
        \int \varphi_2 \varphi_2 - 1
        ,
        \int \varphi_1 \varphi_2
    \right)^T
\]
acting in
\(
    L^2 \left( \mathbb R^3 \right) ^2 \times \mathbb R^4
    .
\)
We are interested in those roots of $\mathcal R$,
points $w$ satisfying \eqref{general_stationary_equation},
which give the lowest possible energy.
At each Newton step we have to deal with \eqref{general_newton}.
The first two lines in the Newton system \eqref{general_newton} have the form
\begin{equation}
\label{newton_component_2_determinant}
    d \mathcal R_i(w) \delta w
    =
    \partial_{\varphi_1} \mathcal R_i(w) \delta \varphi_1
    +
    \ldots
    +
    \frac {\partial \mathcal R_i } {\partial \varepsilon_{12}} (w) \delta \varepsilon_{12}
    , \quad
    i = 1, 2
    ,
\end{equation}
where the first partial derivative is applied to $\delta \varphi_1$ as a linear operator in $L^2$,
and the last partial derivative is a function multiplied by the scalar $\delta \varepsilon_{12}$.
The differential of first component
\(
    d \mathcal R_1(w) \delta w
\)
consists of
\[
    \partial_{\varphi_1} \mathcal R_1(w) \delta \varphi_1
    =
    c_1^2
    \left(
        h + \frac 1{|x|} * \varphi_1^2
    \right)
    \delta \varphi_1
    +
    2 c_1^2
    \left(
        \frac 1{|x|} * (\delta \varphi_1 \varphi_1)
    \right)
    \varphi_1
    +
    c_1 c_2
    \left(
        \frac 1{|x|} * (\delta \varphi_1 \varphi_2)
    \right)
    \varphi_2
    -
    \varepsilon_{11} \delta \varphi_1
    ,
\]

\[
    \partial_{\varphi_2} \mathcal R_1(w) \delta \varphi_2
    =
    c_1 c_2
    \left(
        \frac 1{|x|} * (\varphi_1 \delta \varphi_2)
    \right)
    \varphi_2
    +
    c_1 c_2
    \left(
        \frac 1{|x|} * (\varphi_1 \varphi_2)
    \right)
    \delta \varphi_2
    -
    \varepsilon_{12} \delta \varphi_2
    ,
\]

\[
    \frac {\partial \mathcal R_1 } {\partial \theta} (w) \delta \theta
    =
    - \delta \theta \sin 2\theta
    \left(
        h + \frac 1{|x|} * \varphi_1^2
    \right)
    \varphi_1
    +
    \delta \theta  
    \cos 2 \theta
    \left(
        \frac 1{|x|} * (\varphi_1 \varphi_2)
    \right)
    \varphi_2
\]

and the rest
\[
    \frac {\partial \mathcal R_1 } {\partial \varepsilon_{11}} (w) \delta \varepsilon_{11}
    +
    \frac {\partial \mathcal R_1 } {\partial \varepsilon_{22}} (w) \delta \varepsilon_{22}
    +
    \frac {\partial \mathcal R_1 } {\partial \varepsilon_{12}} (w) \delta \varepsilon_{12}
    =
    -
    \delta \varepsilon_{11} \varphi_1
    -
    \delta \varepsilon_{12} \varphi_2
\]

Similarly,
\(
    d \mathcal R_2(w) \delta w
\)
consists of
\[
    \partial_{\varphi_1} \mathcal R_2(w) \delta \varphi_1
    =
    c_1 c_2
    \left(
        \frac 1{|x|} * (\delta \varphi_1 \varphi_2)
    \right)
    \varphi_1
    +
    c_1 c_2
    \left(
        \frac 1{|x|} * (\varphi_1 \varphi_2)
    \right)
    \delta \varphi_1
    -
    \varepsilon_{12} \delta \varphi_1
    ,
\]
\[
    \partial_{\varphi_2} \mathcal R_2(w) \delta \varphi_2
    =
    c_2^2
    \left(
        h + \frac 1{|x|} * \varphi_2^2
    \right)
    \delta \varphi_2
    +
    2 c_2^2
    \left(
        \frac 1{|x|} * (\delta \varphi_2 \varphi_2)
    \right)
    \varphi_2
    +
    c_1 c_2
    \left(
        \frac 1{|x|} * (\varphi_1 \delta \varphi_2)
    \right)
    \varphi_1
    -
    \varepsilon_{22} \delta \varphi_2
    ,
\]
\[
    \frac {\partial \mathcal R_2 } {\partial \theta} (w) \delta \theta
    =
    \delta \theta \sin 2\theta
    \left(
        h + \frac 1{|x|} * \varphi_2^2
    \right)
    \varphi_2
    +
    \delta \theta  
    \cos 2 \theta
    \left(
        \frac 1{|x|} * (\varphi_1 \varphi_2)
    \right)
    \varphi_1
\]
and the remaining terms
\[
    \frac {\partial \mathcal R_2 } {\partial \varepsilon_{11}} (w) \delta \varepsilon_{11}
    +
    \frac {\partial \mathcal R_2 } {\partial \varepsilon_{22}} (w) \delta \varepsilon_{22}
    +
    \frac {\partial \mathcal R_2 } {\partial \varepsilon_{12}} (w) \delta \varepsilon_{12}
    =
    -
    \delta \varepsilon_{12} \varphi_1
    -
    \delta \varepsilon_{22} \varphi_2
    .
\]

The differential of third component
\(
    d \mathcal R_3(w) \delta w
\)
consists of
\[
    \partial_{\varphi_1} \mathcal R_3(w) \delta \varphi_1
    =
    - 4 \sin 2 \theta
    ( (\delta 1 |h| 1) + (\delta 11|11) )
    +
    4 \cos 2\theta
    (\delta 12|12)
    ,
\]
\[
    \partial_{\varphi_2} \mathcal R_3(w) \delta \varphi_2
    =
    4 \sin 2 \theta
    ( (\delta 2 |h| 2) + (\delta 22|22) )
    +
    4 \cos 2\theta
    (1\delta 2|12)
\]
and
\[
    \frac {\partial \mathcal R_3 } {\partial \theta} (w) \delta \theta
    =
    (
        - 2 \cos 2 \theta
        ( 2(1 |h| 1) + (11|11) )
        +
        2 \cos 2 \theta
        ( 2(2 |h| 2) + (22|22) )
        -
        4 \sin 2\theta
        (12|12)
    )
    \delta \theta
    .
\]

Finally, we have
\[
    d \mathcal R_4(w) \delta w
    =
    \partial_{\varphi_1} \mathcal R_4(w) \delta \varphi_1
    =
    2 \int \varphi_1 \delta \varphi_1
    ,
\]
\[
    d \mathcal R_5(w) \delta w
    =
    \partial_{\varphi_2} \mathcal R_5(w) \delta \varphi_2
    =
    2 \int \varphi_2 \delta \varphi_2
\]
and
\[
    d \mathcal R_6(w) \delta w
    =
    \partial_{\varphi_1} \mathcal R_6(w) \delta \varphi_1
    +
    \partial_{\varphi_2} \mathcal R_6(w) \delta \varphi_2
    =
    \int \delta \varphi_1 \varphi_2
    +
    \int \varphi_1 \delta \varphi_2
    .
\]

\subsection{Self-consistent form}

We reduce the full Newton system to a linear system in
orbital updates
\(
    (\delta \varphi_1, \delta \varphi_2)
\)
by eliminating $\delta \theta$ and $\delta \varepsilon$.
Firstly,
$\delta \theta$ is defined
as
\(
    \delta \theta
    =
    \Theta(\delta \varphi_1, \delta \varphi_2, w)
\)
by
\[
    \delta \theta
    =
    -
    \left(
        \frac {\partial \mathcal R_3 } {\partial \theta} (w)
    \right) ^{-1}
    \left(
        \partial_{\varphi_1} \mathcal R_3(w) \delta \varphi_1
        +
        \partial_{\varphi_2} \mathcal R_3(w) \delta \varphi_2
        +
        \mathcal R_3(w)
    \right)
    .
\]

Secondly,
$\delta \varepsilon_{ij}$ are defined
by
\(
    \delta \varepsilon_{ij}
    =
    \mathcal E_{ij}(\delta \varphi_1, \delta \varphi_2, \delta \theta, w)
\)
that we introduce as follows.
We rewrite the first two lines of the Newton's system as
\begin{equation}
\begin{aligned}
    \delta \varepsilon_{11} \varphi_1
    +
    \delta \varepsilon_{12} \varphi_2
    +
    \varepsilon_{11} \delta \varphi_1
    +
    \varepsilon_{12} \delta \varphi_2
    =
    F(\delta \varphi_1, \delta \varphi_2, \delta \theta, w)
    ,
    \\
    \delta \varepsilon_{12} \varphi_1
    +
    \delta \varepsilon_{22} \varphi_2
    +
    \varepsilon_{12} \delta \varphi_1
    +
    \varepsilon_{22} \delta \varphi_2
    =
    G(\delta \varphi_1, \delta \varphi_2, \delta \theta, w)
    .
\end{aligned}
\end{equation}
Multiplying and integrating these equations by $\varphi_1$ and $\varphi_2$,
we obtain four equations in $\mathbb R$.
Then we exclude $\int \varphi_1 \delta \varphi_1$ and $\int \varphi_2 \delta \varphi_2$
using the fourth and the fifth Newton equations.
Finally, complementing these four equations by the sixth Newton equation
we obtain the following system
\begin{equation}
\label{energy_update_system}
    \begin{pmatrix}
        \norm{\varphi_1}^2
        &
        0
        &
        \int \varphi_1 \varphi_2
        &
        0
        &
        \varepsilon_{12}
        \\
        0
        &
        \norm{\varphi_2}^2
        &
        \int \varphi_1 \varphi_2
        &
        \varepsilon_{12}
        &
        0
        \\
        \int \varphi_1 \varphi_2
        &
        0
        &
        \norm{\varphi_2}^2
        &
        \varepsilon_{11}
        &
        0
        \\
        0
        &
        \int \varphi_1 \varphi_2
        &
        \norm{\varphi_1}^2
        &
        0
        &
        \varepsilon_{22}
        \\
        0
        &
        0
        &
        0
        &
        1
        &
        1
    \end{pmatrix}
    \begin{pmatrix}
        \delta \varepsilon_{11}
        \\
        \delta \varepsilon_{22}
        \\
        \delta \varepsilon_{12}
        \\
        \int \delta \varphi_1 \varphi_2
        \\
        \int \varphi_1 \delta \varphi_2
    \end{pmatrix}
    =
    \begin{pmatrix}
        \int F \varphi_1 - \frac 12 \varepsilon_{11} \left( 1 - \norm{\varphi_1}^2 \right)
        \\
        \int G \varphi_2 - \frac 12 \varepsilon_{22} \left( 1 - \norm{\varphi_2}^2 \right)
        \\
        \int F \varphi_2 - \frac 12 \varepsilon_{12} \left( 1 - \norm{\varphi_2}^2 \right)
        \\
        \int G \varphi_1 - \frac 12 \varepsilon_{12} \left( 1 - \norm{\varphi_1}^2 \right)
        \\
        - \int \varphi_1 \varphi_2
    \end{pmatrix}
\end{equation}
We search for orthonormal orbitals $\varphi_1, \varphi_2$.
Therefore, the determinant of this system
should stay
close to $- \varepsilon_{11} - \varepsilon_{22}$ during each Newton's update.
In particular, it ensures that the matrix
is invertible for each Newton's iteration $w = w_n$.
The integrals staying in the right hand side of \eqref{energy_update_system}
are computed as follows
\begin{multline}
\label{integral_F_phi_1}
    \int F \varphi_1
    =
    c_1^2 ( \delta 1 |h| 1 )
    +
    3 c_1^2 ( \delta 11 | 11 )
    +
    c_1 c_2 ( \delta 12 | 12 )
    +
    2 c_1 c_2 ( 1 \delta 2 | 12 )
    \\
    +
    (
        c_1^2 - \delta \theta \sin 2\theta
    )
    (
        ( 1 |h| 1 ) + ( 11 | 11 )
    )
    +
    (
        c_1 c_2 + \delta \theta \cos 2\theta
    )
    ( 12 | 12 )
    -
    \varepsilon_{11} \norm{\varphi_1}^2
    -
    \varepsilon_{12} \int \varphi_1 \varphi_2
    ,
\end{multline}
\begin{multline}
\label{integral_F_phi_2}
    \int F \varphi_2
    =
    c_1^2 ( \delta 1 |h| 2 )
    +
    c_1^2 ( \delta 12 | 11 )
    +
    2 c_1^2 ( \delta 11 | 12 )
    +
    c_1 c_2 ( \delta 12 | 22 )
    +
    c_1 c_2 ( 1 \delta 2 | 22 )
    +
    c_1 c_2 ( 2 \delta 2 | 12 )
    \\
    +
    (
        c_1^2 - \delta \theta \sin 2\theta
    )
    (
        ( 1 |h| 2 ) + ( 11 | 12 )
    )
    +
    (
        c_1 c_2 + \delta \theta \cos 2\theta
    )
    ( 12 | 22 )
    -
    \varepsilon_{11} \int \varphi_1 \varphi_2
    -
    \varepsilon_{12} \norm{\varphi_2}^2
    ,
\end{multline}
\begin{multline}
\label{integral_G_phi_2}
    \int G \varphi_2
    =
    c_2^2 ( \delta 2 |h| 2 )
    +
    3 c_2^2 ( \delta 22 | 22 )
    +
    2 c_1 c_2 ( \delta 12 | 12 )
    +
    c_1 c_2 ( 1 \delta 2 | 12 )
    \\
    +
    (
        c_2^2 + \delta \theta \sin 2\theta
    )
    (
        ( 2 |h| 2 ) + ( 22 | 22 )
    )
    +
    (
        c_1 c_2 + \delta \theta \cos 2\theta
    )
    ( 12 | 12 )
    -
    \varepsilon_{22} \norm{\varphi_2}^2
    -
    \varepsilon_{12} \int \varphi_1 \varphi_2
    ,
\end{multline}
\begin{multline}
\label{integral_G_phi_1}
    \int G \varphi_1
    =
    c_2^2 ( \delta 2 |h| 1 )
    +
    c_2^2 ( 1 \delta 2 | 22 )
    +
    2 c_2^2 ( \delta 22 | 12 )
    +
    c_1 c_2 ( \delta 12 | 11 )
    +
    c_1 c_2 ( 1 \delta 2 | 11 )
    +
    c_1 c_2 ( 1 \delta 1 | 12 )
    \\
    +
    (
        c_2^2 + \delta \theta \sin 2\theta
    )
    (
        ( 1 |h| 2 ) + ( 12 | 22 )
    )
    +
    (
        c_1 c_2 + \delta \theta \cos 2\theta
    )
    ( 11 | 12 )
    -
    \varepsilon_{22} \int \varphi_1 \varphi_2
    -
    \varepsilon_{12} \norm{\varphi_1}^2
    .
\end{multline}

It is left to rearrange the functional equations
\begin{multline}
    c_1^2
    \left(
        h + \frac 1{|x|} * \varphi_1^2
    \right)
    \delta \varphi_1
    +
    2 c_1^2
    \left(
        \frac 1{|x|} * (\delta \varphi_1 \varphi_1)
    \right)
    \varphi_1
    +
    c_1 c_2
    \left(
        \frac 1{|x|} * (\delta \varphi_1 \varphi_2)
    \right)
    \varphi_2
    -
    \varepsilon_{11} \delta \varphi_1
    \\
    +
    c_1 c_2
    \left(
        \frac 1{|x|} * (\varphi_1 \delta \varphi_2)
    \right)
    \varphi_2
    +
    c_1 c_2
    \left(
        \frac 1{|x|} * (\varphi_1 \varphi_2)
    \right)
    \delta \varphi_2
    -
    \varepsilon_{12} \delta \varphi_2
    \\
    - \delta \theta \sin 2\theta
    \left(
        h + \frac 1{|x|} * \varphi_1^2
    \right)
    \varphi_1
    +
    \delta \theta  
    \cos 2 \theta
    \left(
        \frac 1{|x|} * (\varphi_1 \varphi_2)
    \right)
    \varphi_2
    -
    \delta \varepsilon_{11} \varphi_1
    -
    \delta \varepsilon_{12} \varphi_2
    \\
    +
    c_1^2
    \left(
        h + \frac 1{|x|} * \varphi_1^2
    \right)
    \varphi_1
    +
    c_1 c_2
    \left(
        \frac 1{|x|} * (\varphi_1 \varphi_2)
    \right)
    \varphi_2
    -
    \varepsilon_{11} \varphi_1
    -
    \varepsilon_{12} \varphi_2
    =
    0
\end{multline}
and
\begin{multline}
    c_1 c_2
    \left(
        \frac 1{|x|} * (\delta \varphi_1 \varphi_2)
    \right)
    \varphi_1
    +
    c_1 c_2
    \left(
        \frac 1{|x|} * (\varphi_1 \varphi_2)
    \right)
    \delta \varphi_1
    -
    \varepsilon_{12} \delta \varphi_1
    \\
    +
    c_2^2
    \left(
        h + \frac 1{|x|} * \varphi_2^2
    \right)
    \delta \varphi_2
    +
    2 c_2^2
    \left(
        \frac 1{|x|} * (\delta \varphi_2 \varphi_2)
    \right)
    \varphi_2
    +
    c_1 c_2
    \left(
        \frac 1{|x|} * (\varphi_1 \delta \varphi_2)
    \right)
    \varphi_1
    -
    \varepsilon_{22} \delta \varphi_2
    \\
    +
    \delta \theta \sin 2\theta
    \left(
        h + \frac 1{|x|} * \varphi_2^2
    \right)
    \varphi_2
    +
    \delta \theta  
    \cos 2 \theta
    \left(
        \frac 1{|x|} * (\varphi_1 \varphi_2)
    \right)
    \varphi_1
    -
    \delta \varepsilon_{12} \varphi_1
    -
    \delta \varepsilon_{22} \varphi_2
    \\
    +
    c_2^2
    \left(
        h + \frac 1{|x|} * \varphi_2^2
    \right)
    \varphi_2
    +
    c_1 c_2
    \left(
        \frac 1{|x|} * (\varphi_1 \varphi_2)
    \right)
    \varphi_1
    -
    \varepsilon_{12} \varphi_1
    -
    \varepsilon_{22} \varphi_2
    =
    0
    .
\end{multline}
Using the short notation $J$ for convolutions,
these two equations can rewritten as
\begin{multline}
    c_1^2
    \left(
        h + J (11)
    \right)
    \delta \varphi_1
    +
    2 c_1^2
    J (1 \delta 1)
    \varphi_1
    +
    c_1 c_2
    J (\delta 12)
    \varphi_2
    -
    \varepsilon_{11} \delta \varphi_1
    \\
    +
    c_1 c_2
    J (1 \delta 2)
    \varphi_2
    +
    c_1 c_2
    J (12)
    \delta \varphi_2
    -
    \varepsilon_{12} \delta \varphi_2
    \\
    - \delta \theta \sin 2\theta
    \left(
        h + J(11)
    \right)
    \varphi_1
    +
    \delta \theta  
    \cos 2 \theta
    J (12)
    \varphi_2
    -
    \delta \varepsilon_{11} \varphi_1
    -
    \delta \varepsilon_{12} \varphi_2
    \\
    +
    c_1^2
    \left(
        h + J(11)
    \right)
    \varphi_1
    +
    c_1 c_2
    J (12)
    \varphi_2
    -
    \varepsilon_{11} \varphi_1
    -
    \varepsilon_{12} \varphi_2
    =
    0
\end{multline}
and
\begin{multline}
    c_1 c_2
    J (\delta 12)
    \varphi_1
    +
    c_1 c_2
    J (12)
    \delta \varphi_1
    -
    \varepsilon_{12} \delta \varphi_1
    \\
    +
    c_2^2
    \left(
        h + J(22)
    \right)
    \delta \varphi_2
    +
    2 c_2^2
    J (2 \delta 2)
    \varphi_2
    +
    c_1 c_2
    J (1 \delta 2)
    \varphi_1
    -
    \varepsilon_{22} \delta \varphi_2
    \\
    +
    \delta \theta \sin 2\theta
    \left(
        h + J(22)
    \right)
    \varphi_2
    +
    \delta \theta  
    \cos 2 \theta
    J (12)
    \varphi_1
    -
    \delta \varepsilon_{12} \varphi_1
    -
    \delta \varepsilon_{22} \varphi_2
    \\
    +
    c_2^2
    \left(
        h
        + J(22)
    \right)
    \varphi_2
    +
    c_1 c_2
    J (12)
    \varphi_1
    -
    \varepsilon_{12} \varphi_1
    -
    \varepsilon_{22} \varphi_2
    =
    0
    .
\end{multline}
Equivalently,
\begin{multline}
    c_1^2
    \left(
        - \frac 12 \Delta + V_{\text{nuc}}
        +
        J (11)
    \right)
    \delta \varphi_1
    +
    2 c_1^2
    J (1 \delta 1)
    \varphi_1
    +
    c_1 c_2
    J (\delta 12)
    \varphi_2
    -
    \varepsilon_{11} \delta \varphi_1
    \\
    +
    c_1 c_2
    J (1 \delta 2)
    \varphi_2
    +
    c_1 c_2
    J (12)
    \delta \varphi_2
    -
    \varepsilon_{12} \delta \varphi_2
    \\
    -
    \delta \varepsilon_{11} \varphi_1
    -
    \delta \varepsilon_{12} \varphi_2
    -
    \varepsilon_{11} \varphi_1
    -
    \varepsilon_{12} \varphi_2
    \\
    +
    \left(
        c_1^2
        - \delta \theta \sin 2\theta
    \right)
    \left(
        - \frac 12 \Delta + V_{\text{nuc}}
        +
        J(11)
    \right)
    \varphi_1
    +
    (    
        c_1 c_2
        +
        \delta \theta  
        \cos 2 \theta
    )
    J (12)
    \varphi_2
    =
    0
\end{multline}
and
\begin{multline}
    c_1 c_2
    J (\delta 12)
    \varphi_1
    +
    c_1 c_2
    J (12)
    \delta \varphi_1
    -
    \varepsilon_{12} \delta \varphi_1
    \\
    +
    c_2^2
    \left(
        - \frac 12 \Delta + V_{\text{nuc}}
        +
        J(22)
    \right)
    \delta \varphi_2
    +
    2 c_2^2
    J (2 \delta 2)
    \varphi_2
    +
    c_1 c_2
    J (1 \delta 2)
    \varphi_1
    -
    \varepsilon_{22} \delta \varphi_2
    \\
    -
    \delta \varepsilon_{12} \varphi_1
    -
    \delta \varepsilon_{22} \varphi_2
    -
    \varepsilon_{12} \varphi_1
    -
    \varepsilon_{22} \varphi_2
    \\
    +
    \left(
        c_2^2
        +
        \delta \theta \sin 2\theta
    \right)
    \left(
        - \frac 12 \Delta + V_{\text{nuc}}
        +
        J(22)
    \right)
    \varphi_2
    +
    (
        c_1 c_2
        +
        \delta \theta  
        \cos 2 \theta
    )
    J (12)
    \varphi_1
    =
    0
    .
\end{multline}
Finally, we rewrite these two equations in the self consistent form,
by first regrupping the terms as follows
\begin{multline}
    \left(
        - \frac {c_1^2}2 \Delta
        -
        \varepsilon_{11}
    \right)
    \delta \varphi_1
    +
    c_1^2
    \left(
        V_{\text{nuc}}
        +
        J (11)
    \right)
    \delta \varphi_1
    \\
    +
    2 c_1^2
    J (1 \delta 1)
    \varphi_1
    +
    c_1 c_2
    J (\delta 12)
    \varphi_2
    +
    c_1 c_2
    J (1 \delta 2)
    \varphi_2
    +
    c_1 c_2
    J (12)
    \delta \varphi_2
    -
    \varepsilon_{12} \delta \varphi_2
    \\
    -
    \delta \varepsilon_{11} \varphi_1
    -
    \delta \varepsilon_{12} \varphi_2
    -
    \varepsilon_{11} \varphi_1
    -
    \varepsilon_{12} \varphi_2
    \\
    +
    \left(
        1
        -
        \frac {\delta \theta \sin 2\theta}{c_1^2}
    \right)
    \left(
        - \frac {c_1^2}2 \Delta
    \right)
    \varphi_1
    +
    \left(
        c_1^2
        - \delta \theta \sin 2\theta
    \right)
    \left(
        V_{\text{nuc}}
        +
        J(11)
    \right)
    \varphi_1
    +
    (    
        c_1 c_2
        +
        \delta \theta  
        \cos 2 \theta
    )
    J (12)
    \varphi_2
    =
    0
    ,
\end{multline}
\begin{multline}
    c_1 c_2
    J (\delta 12)
    \varphi_1
    +
    c_1 c_2
    J (12)
    \delta \varphi_1
    +
    2 c_2^2
    J (2 \delta 2)
    \varphi_2
    +
    c_1 c_2
    J (1 \delta 2)
    \varphi_1
    -
    \varepsilon_{12} \delta \varphi_1
    \\
    +
    \left(
        -
        \frac {c_2^2}2 \Delta
        -
        \varepsilon_{22}
    \right)
    \delta \varphi_2
    +
    c_2^2
    \left(
        V_{\text{nuc}}
        +
        J(22)
    \right)
    \delta \varphi_2
    \\
    -
    \delta \varepsilon_{12} \varphi_1
    -
    \delta \varepsilon_{22} \varphi_2
    -
    \varepsilon_{12} \varphi_1
    -
    \varepsilon_{22} \varphi_2
    \\
    +
    \left(
        1
        +
        \frac {\delta \theta \sin 2\theta}{c_2^2}
    \right)
    \left(
        - \frac {c_2^2}2 \Delta
    \right)
    \varphi_2
    +
    \left(
        c_2^2
        +
        \delta \theta \sin 2\theta
    \right)
    \left(
        V_{\text{nuc}}
        +
        J(22)
    \right)
    \varphi_2
    +
    (
        c_1 c_2
        +
        \delta \theta  
        \cos 2 \theta
    )
    J (12)
    \varphi_1
    =
    0
    .
\end{multline}
Eventually,
we obtain two equations,
where the kinetic part has the pattern
\(
    - \frac {c_i^2}{2} \Delta - \varepsilon_{ii}
    ,
\)
that can be easily inverted.
Indeed,
we have
\begin{multline}
    \left(
        - \frac {c_1^2}2 \Delta
        -
        \varepsilon_{11}
    \right)
    \delta \varphi_1
    +
    \left(
        1
        -
        \frac {\delta \theta \sin 2\theta}{c_1^2}
    \right)
    \left(
        - \frac {c_1^2}2 \Delta
        - \varepsilon_{11}
    \right)
    \varphi_1
    \\
    +
    c_1^2
    \left(
        V_{\text{nuc}}
        +
        J (11)
    \right)
    \delta \varphi_1
    +
    2 c_1^2
    J (1 \delta 1)
    \varphi_1
    +
    c_1 c_2
    J (\delta 12)
    \varphi_2
    +
    c_1 c_2
    J (1 \delta 2)
    \varphi_2
    +
    c_1 c_2
    J (12)
    \delta \varphi_2
    -
    \varepsilon_{12} \delta \varphi_2
    \\
    -
    \delta \varepsilon_{12} \varphi_2
    -
    \varepsilon_{12} \varphi_2
    -
    \left(
        \varepsilon_{11} \frac {\delta \theta \sin 2\theta}{c_1^2}
        +
        \delta \varepsilon_{11}
    \right)
    \varphi_1
    \\
    +
    \left(
        c_1^2
        - \delta \theta \sin 2\theta
    \right)
    \left(
        V_{\text{nuc}}
        +
        J(11)
    \right)
    \varphi_1
    +
    (    
        c_1 c_2
        +
        \delta \theta  
        \cos 2 \theta
    )
    J (12)
    \varphi_2
    =
    0
\end{multline}
and
\begin{multline}
    \left(
        -
        \frac {c_2^2}2 \Delta
        -
        \varepsilon_{22}
    \right)
    \delta \varphi_2
    +
    \left(
        1
        +
        \frac {\delta \theta \sin 2\theta}{c_2^2}
    \right)
    \left(
        - \frac {c_2^2}2 \Delta
        - \varepsilon_{22}
    \right)
    \varphi_2
    \\
    +
    c_1 c_2
    J (\delta 12)
    \varphi_1
    +
    c_1 c_2
    J (12)
    \delta \varphi_1
    +
    2 c_2^2
    J (2 \delta 2)
    \varphi_2
    +
    c_1 c_2
    J (1 \delta 2)
    \varphi_1
    -
    \varepsilon_{12} \delta \varphi_1
    +
    c_2^2
    \left(
        V_{\text{nuc}}
        +
        J(22)
    \right)
    \delta \varphi_2
    \\
    -
    \delta \varepsilon_{12} \varphi_1
    -
    \varepsilon_{12} \varphi_1
    +
    \left(    
        \varepsilon_{22} \frac {\delta \theta \sin 2\theta}{c_2^2}
        -
        \delta \varepsilon_{22}
    \right)    
    \varphi_2
    \\
    +
    \left(
        c_2^2
        +
        \delta \theta \sin 2\theta
    \right)
    \left(
        V_{\text{nuc}}
        +
        J(22)
    \right)
    \varphi_2
    +
    (
        c_1 c_2
        +
        \delta \theta  
        \cos 2 \theta
    )
    J (12)
    \varphi_1
    =
    0
    .
\end{multline}

Finally,
inverting the kinetic energy
we obtain the system
\begin{equation}
\begin{aligned}
    \delta \varphi_1
    =
    -
    \left(
        1
        -
        \frac {\delta \theta \sin 2\theta}{c_1^2}
    \right)
    \varphi_1
    -
    R_1 \mathfrak F_1
    ,
    \\
    \delta \varphi_2
    =
    -
    \left(
        1
        +
        \frac {\delta \theta \sin 2\theta}{c_2^2}
    \right)
    \varphi_2
    -
    R_2 \mathfrak F_2
    ,
\end{aligned}
\end{equation}
where
\begin{multline*}
    \mathfrak F_1
    =
    -
    \varepsilon_{12} \delta \varphi_2
    -
    (
        \delta \varepsilon_{12}
        +
        \varepsilon_{12}
    )
    \varphi_2
    -
    \left(
        \varepsilon_{11} \frac {\delta \theta \sin 2\theta}{c_1^2}
        +
        \delta \varepsilon_{11}
    \right)
    \varphi_1
    \\
    +
    \left(
        V_{\text{nuc}}
        +
        J (11)
    \right)
    \left(
        c_1^2
        \delta \varphi_1
        +
        \left(
            c_1^2
            - \delta \theta \sin 2\theta
        \right)
        \varphi_1
    \right)
    +
    J (12)
    (
        c_1 c_2
        \delta \varphi_2
        +
        (    
            c_1 c_2
            +
            \delta \theta  
            \cos 2 \theta
        )
        \varphi_2
    )
    \\
    +
    2 c_1^2
    J (1 \delta 1)
    \varphi_1
    +
    c_1 c_2
    J (\delta 12)
    \varphi_2
    +
    c_1 c_2
    J (1 \delta 2)
    \varphi_2
\end{multline*}
and
\begin{multline*}
    \mathfrak F_2
    =
    -
    \varepsilon_{12} \delta \varphi_1
    -
    (
        \delta \varepsilon_{12}
        +
        \varepsilon_{12}
    )
    \varphi_1
    +
    \left(    
        \varepsilon_{22} \frac {\delta \theta \sin 2\theta}{c_2^2}
        -
        \delta \varepsilon_{22}
    \right)    
    \varphi_2
    \\
    +
    \left(
        V_{\text{nuc}}
        +
        J(22)
    \right)
    \left(
        c_2^2
        \delta \varphi_2
        +
        \left(
            c_2^2
            +
            \delta \theta \sin 2\theta
        \right)
        \varphi_2
    \right)
    +
    J (12)
    (
        c_1 c_2
        \delta \varphi_1
        +
        (
            c_1 c_2
            +
            \delta \theta  
            \cos 2 \theta
        )
        \varphi_1
    )
    \\
    +
    2 c_2^2
    J (2 \delta 2)
    \varphi_2
    +
    c_1 c_2
    J (\delta 12)
    \varphi_1
    +
    c_1 c_2
    J (1 \delta 2)
    \varphi_1
    .
\end{multline*}
This completes the description of the self-consistent form \eqref{general_self_consistent_form},
suitable for numerical treatment of the Newton equation \eqref{general_newton}
in the two-determinant case.
These equations can be easily discretized 
using multiwavelets.

\begin{figure}[ht!]
    \centering
    {
        \begin{tikzpicture}[scale=3, >=Stealth]
        
            \pgfmathsetmacro{\xA}{0.3 * pi}
            \pgfmathsetmacro{\yA}{cos(deg(\xA))}
            
            \pgfmathsetmacro{\xB}{\xA + 1.9}
            \pgfmathsetmacro{\yB}{\yA + 0.0}
            
            \pgfmathsetmacro{\xC}{0.61 * pi}
            \pgfmathsetmacro{\yC}{cos(deg(\xC))}
            
            \pgfmathsetmacro{\xD}{1.11 * pi}
            \pgfmathsetmacro{\yD}{cos(deg(\xD))}
            
            \draw[thick, domain = 0.05 * pi : 1.2 * pi, smooth, variable=\x]
                plot ({\x}, {cos(deg(\x))})
                node[above left] at (0.5, 1.0) {constrain surface};
            
            \coordinate (A) at (\xA, \yA);
            \coordinate (B) at (\xB, \yB);
            \coordinate (C) at (\xC, \yC);
            \coordinate (D) at (\xD, \yD);
            
            \fill (A) circle (0.8pt);
            \fill (B) circle (0.8pt);
            \fill (C) circle (0.8pt);
            \fill (D) circle (0.8pt);
            
            \node[below left] at (A) {$\left( \varphi^{(n)}, c^{(n)} \right)$};
              
            \draw[->, thick] (A) -- (B) node[midway, sloped, above] {Newton};
            \draw[->, thick] (B) -- (C) node[midway, sloped, above] {L\"owdin};
            \draw[->, thick] (C) -- (D) node[midway, sloped, above] {$\varphi = \text{const}$};
            
            \node[below right] at (D) {$\left( \varphi^{(n+1)}, c^{(n+1)} \right)$};
        
        \end{tikzpicture}
    }
    \caption
    {
	Newton energy optimization step.
    }
\label{Newton_optimisation_figure}
\end{figure}

\begin{figure}[ht!]
    \centering
    {
	\includegraphics[width=0.7\textwidth]
        {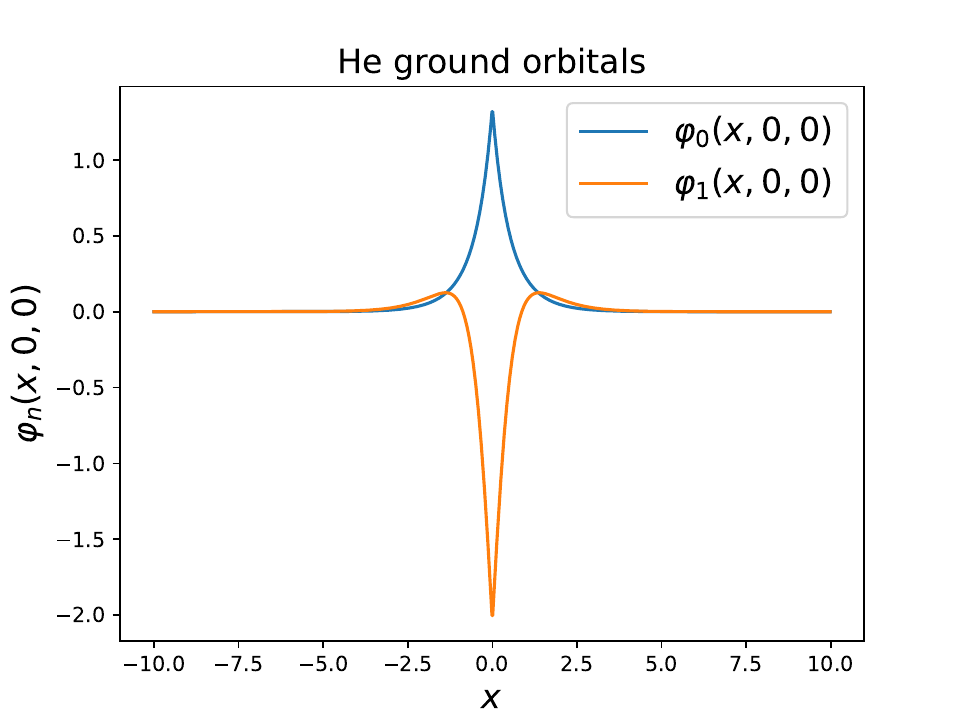}
    }
    \caption
    {
    	Two-determinant ground state of the helium atom
        with the coefficients $c_0 = 0.99793$ and $ c_1 = -0.06430$, correspondingly.
        The corresponding total energy is $-2.87799$.
    }
\label{ground_helium_orbitals_figure}
\end{figure}

Iterative Newton procedure fits into two loops.
The inner loop is associated with solving Newton's equation at a given Newton step.
One can use simple iteration or \ac{DIIS} acceleration over unknown $x = (\delta \varphi_1, \delta \varphi_2)$ as
\[
    \delta \varphi_1, \delta \varphi_2 = 0
    \mapsto
    \delta \theta
    \mapsto
    \delta \varepsilon_{ij}
    \mapsto
    \delta \varphi_1, \delta \varphi_2
    \mapsto
    \ldots
    .
\]
It is terminated when either $\norm{x_{n+1} - x_n}$
or the residual is smaller than a given tolerance, depending on $\varepsilon_{\text{mra}}$
for multiwavelet discretization.
The outer loop corresponds to the Newton step
\(
    w \mapsto w + \delta w
    .
\)
Newton's method is sensible to the choice $w_0$ of an initial guess.
Therefore, a trust radius technique or a level shifting (described in the next section)
should guide each iteration $w_n \mapsto w_{n + 1}$,
which does not lead to the decrease of energy.
Alternatively, for the first couple of iterations one can intervene to the Newton's procedure,
in order to direct it in the right way:
by setting $\varepsilon_{22}^{n+1} = 10 * (\varepsilon_{22}^n + \delta \varepsilon_{22})$.
This trick is related to the level-shifting.
At the beginning orbital energy matrix is initialized by $\varepsilon_{ij}^0 = - \delta_{ij}$.
The orbitals of an initial point $w_0$ can be formed from $1s, 2s$ orbitals
of the corresponding ion, He$^+$ or H2$^+$, for example.
Initial guess for $\theta \in (- \pi / 4, 0) \cup (0, \pi / 4)$ can be taken from
\(
    \frac{\partial E}{\partial \theta} (\varphi_1, \varphi_2, \theta) = 0
    ,
\)
provided $\varphi_1, \varphi_2$ are given, as follows
\[
    \tan 2 \theta
    =
    \frac
    {
        2 (12|12)
    }
    {
        2(1 |h| 1) + (11|11) - 2(2 |h| 2) - (22|22)
    }
    .
\]
Overall, the complete Newton procedure can be
schematically described by Figure \ref{Newton_optimisation_figure}.
It is not necessary to perform the Löwdin transformation and the optimization of CI coefficients
after every Newton iteration;
however, including these steps significantly accelerates convergence.
In other words, the dominant factor affecting convergence is the treatment of normalization.
To achieve second-order convergence,
the orbitals must be normalized after each outer-loop iteration.
The two orbital \ac{MCSCF} for the helium atom results in the coefficients
$c_1 = 0.99793, c_2 = -0.06430$, associated to $\theta = -0.06435$,
and orbitals symmetric with respect to origin,
shown in Figure \ref{ground_helium_orbitals_figure}.
The corresponding total energy is $-2.87799$.

\section{Symmetric hessian and shift}
\setcounter{equation}{0}

In the previous section we did not bother to define $\mathcal R(w)$
so the operator $d \mathcal R(w)$ is symmetric.
However, the hermitian property of the second derivative is important
for introducing the Levenberg–Marquardt damping $\lambda \geqslant 0$.
The Lagrangian is defined by
\[
    \mathcal L
    =
    \frac 14 E - \frac 12 \sum_{i, j  = 1}^2 \varepsilon_{ij} \left( \int \varphi_i \varphi_j - \delta_{ij} \right)
\]
with symmetric matrix $\varepsilon$ and variational derivatives
\begin{equation}
\begin{aligned}
    \frac{\delta \mathcal L}{\delta \varphi_1}
    =
    c_1^2
    \left(
        h + \frac 1{|x|} * \varphi_1^2
    \right)
    \varphi_1
    +
    c_1 c_2
    \left(
        \frac 1{|x|} * (\varphi_1 \varphi_2)
    \right)
    \varphi_2
    -
    \varepsilon_{11} \varphi_1
    -
    \varepsilon_{12} \varphi_2
    ,
    \\
    \frac{\delta \mathcal L}{\delta \varphi_2}
    =
    c_2^2
    \left(
        h + \frac 1{|x|} * \varphi_2^2
    \right)
    \varphi_2
    +
    c_1 c_2
    \left(
        \frac 1{|x|} * (\varphi_1 \varphi_2)
    \right)
    \varphi_1
    -
    \varepsilon_{12} \varphi_1
    -
    \varepsilon_{22} \varphi_2
    ,
\end{aligned}
\end{equation}
where $c_1 = \cos \theta$ and $c_2 = \sin \theta$.
The derivatives with respect to scalars are
\[
    \frac{\partial \mathcal L}{\partial \theta}
    =
    - \frac 14 \sin 2 \theta
    ( 2(1 |h| 1) + (11|11) )
    +
    \frac 14 \sin 2 \theta
    ( 2(2 |h| 2) + (22|22) )
    +
    \frac 12 \cos 2\theta
    (12|12)
\]
and
\begin{equation}
\begin{aligned}
    \frac{\partial \mathcal L}{\partial \varepsilon_{11}}
    &=
    - \frac 12 \left( \norm{\varphi_1}^2 - 1 \right)
    ,
    \\
    \frac{\partial \mathcal L}{\partial \varepsilon_{22}}
    &=
    - \frac 12 \left( \norm{\varphi_2}^2 - 1 \right)
    ,
    \\
    \frac{\partial \mathcal L}{\partial \varepsilon_{12}}
    &=
    - \int \varphi_1 \varphi_2
    .
\end{aligned}
\end{equation}

We introduce
\[
    \mathcal R( \varphi_1, \varphi_2, \theta, \varepsilon_{11}, \varepsilon_{22}, \varepsilon_{12} )
    =
    \nabla \mathcal L
    =
    \left(
        \frac{\delta \mathcal L}{\delta \varphi_1}
        ,
        \frac{\delta \mathcal L}{\delta \varphi_2}
        ,
        \frac{\partial \mathcal L}{\partial \theta}
        ,
        \frac{\partial \mathcal L}{\partial \varepsilon_{11}}
        ,
        \frac{\partial \mathcal L}{\partial \varepsilon_{22}}
        ,
        \frac{\partial \mathcal L}{\partial \varepsilon_{12}}
    \right)^T
\]
acting in
\(
    W
    =
    L^2 \left( \mathbb R^3 \right) ^2 \times \mathbb R^4
    .
\)
Now for a given $w = w_n \in W$ we solve the shifted Newton's equation
\[
    ( d \mathcal R(w) + \lambda )
    \delta w
    =
    - \mathcal R(w)
    , \quad
    \lambda \geqslant 0
    ,
\]
with respect $\delta w \in W$.
This is a linear equation that we solve iteratively
and accelerating with \ac{DIIS}.
The Jacobian $d \mathcal R(w)$ equals to the Hessian $d \nabla \mathcal L(w)$.
As above, the solution of this equation 
allows to update the current iterate $w_n$ by $w_{n + 1} = w_n + \delta w$.
The derivatives of the first two components \eqref{newton_component_2_determinant}
were found in the previous section, 
and they are exactly the same here for the new $\mathcal R$.

The differential of third component
\(
    d \mathcal R_3(w) \delta w
\)
consists of
\[
    \partial_{\varphi_1} \mathcal R_3(w) \delta \varphi_1
    =
    -
    \sin 2 \theta
    ( (\delta 1 |h| 1) + (\delta 11|11) )
    +
    \cos 2\theta
    (\delta 12|12)
    ,
\]
\[
    \partial_{\varphi_2} \mathcal R_3(w) \delta \varphi_2
    =
    \sin 2 \theta
    ( (\delta 2 |h| 2) + (\delta 22|22) )
    +
    \cos 2\theta
    (1\delta 2|12)
\]
and
\[
    \frac {\partial \mathcal R_3 } {\partial \theta} (w) \delta \theta
    =
    \left(
        -
        \frac 12 \cos 2 \theta
        ( 2(1 |h| 1) + (11|11) )
        +
        \frac 12 \cos 2 \theta
        ( 2(2 |h| 2) + (22|22) )
        -
        \sin 2\theta
        (12|12)
    \right)
    \delta \theta
    .
\]

Finally,
\[
    d \mathcal R_4(w) \delta w
    =
    \partial_{\varphi_1} \mathcal R_4(w) \delta \varphi_1
    =
    - \int \varphi_1 \delta \varphi_1
    ,
\]
\[
    d \mathcal R_5(w) \delta w
    =
    \partial_{\varphi_2} \mathcal R_5(w) \delta \varphi_2
    =
    - \int \varphi_2 \delta \varphi_2
    ,
\]
\[
    d \mathcal R_6(w) \delta w
    =
    \partial_{\varphi_1} \mathcal R_6(w) \delta \varphi_1
    +
    \partial_{\varphi_2} \mathcal R_6(w) \delta \varphi_2
    =
    -
    \int \delta \varphi_1 \varphi_2
    -
    \int \varphi_1 \delta \varphi_2
    .
\]
This finishes the detailed description of the Newton equation.
Now we want to rewrite it in the self-consistent form,
which will be suitable for an iterative procedure.

\subsection{Self-consistent form}

We rewrite the Newton equation in the form
\[
    (\delta \varphi_1, \delta \varphi_2)^T
    =
    \mathbb F( \delta \varphi_1, \delta \varphi_2; w )
    ,
\]
where
\(
    w = ( \varphi_1, \varphi_2, \theta, \varepsilon_{11}, \varepsilon_{22}, \varepsilon_{12} )
\)
is fixed at each Newton step,
as follows.
Firstly,
$\delta \theta$ is defined
as
\(
    \delta \theta
    =
    \Theta(\delta \varphi_1, \delta \varphi_2, w)
\)
by
\[
    \delta \theta
    =
    -
    \left(
        \frac {\partial \mathcal R_3 } {\partial \theta} (w)
        +
        \lambda
    \right) ^{-1}
    \left(
        \partial_{\varphi_1} \mathcal R_3(w) \delta \varphi_1
        +
        \partial_{\varphi_2} \mathcal R_3(w) \delta \varphi_2
        +
        \mathcal R_3(w)
    \right)
    .
\]

Secondly,
$\delta \varepsilon_{ij}$ are defined
by
\(
    \delta \varepsilon_{ij}
    =
    \mathcal E_{ij}(\delta \varphi_1, \delta \varphi_2, \delta \theta, w)
\)
that we introduce as follows.
We rewrite the first two lines of the Newton's system as
\begin{equation}
\begin{aligned}
    \delta \varepsilon_{11} \varphi_1
    +
    \delta \varepsilon_{12} \varphi_2
    +
    (\varepsilon_{11} - \lambda) \delta \varphi_1
    +
    \varepsilon_{12} \delta \varphi_2
    =
    F(\delta \varphi_1, \delta \varphi_2, \delta \theta, w)
    ,
    \\
    \delta \varepsilon_{12} \varphi_1
    +
    \delta \varepsilon_{22} \varphi_2
    +
    \varepsilon_{12} \delta \varphi_1
    +
    (\varepsilon_{22} - \lambda) \delta \varphi_2
    =
    G(\delta \varphi_1, \delta \varphi_2, \delta \theta, w)
    .
\end{aligned}
\end{equation}
Multiplying these equations by $\varphi_1$, $\varphi_2$ and then integrating,
we obtain four equations in $\mathbb R$.
We exclude $\int \varphi_1 \delta \varphi_1$ and $\int \varphi_2 \delta \varphi_2$
using the fourth and the fifth Newton equations.
Finally, complementing these four equations by the sixth Newton equation
we obtain the following system

\begin{multline}
\label{energy_update_system_shifted}
    \begin{pmatrix}
        \norm{\varphi_1}^2
        +
        (\varepsilon_{11} - \lambda) \lambda
        &
        0
        &
        \int \varphi_1 \varphi_2
        &
        0
        &
        \varepsilon_{12}
        \\
        0
        &
        \norm{\varphi_2}^2
        +
        (\varepsilon_{22} - \lambda) \lambda
        &
        \int \varphi_1 \varphi_2
        &
        \varepsilon_{12}
        &
        0
        \\
        \int \varphi_1 \varphi_2
        &
        0
        &
        \norm{\varphi_2}^2
        &
        \varepsilon_{11} - \lambda
        &
        0
        \\
        0
        &
        \int \varphi_1 \varphi_2
        &
        \norm{\varphi_1}^2
        &
        0
        &
        \varepsilon_{22} - \lambda
        \\
        0
        &
        0
        &
        - \lambda
        &
        1
        &
        1
    \end{pmatrix}
    \begin{pmatrix}
        \delta \varepsilon_{11}
        \\
        \delta \varepsilon_{22}
        \\
        \delta \varepsilon_{12}
        \\
        \int \delta \varphi_1 \varphi_2
        \\
        \int \varphi_1 \delta \varphi_2
    \end{pmatrix}
    \\
    =
    \begin{pmatrix}
        \int F \varphi_1 - \frac 12 (\varepsilon_{11} - \lambda) \left( 1 - \norm{\varphi_1}^2 \right)
        \\
        \int G \varphi_2 - \frac 12 (\varepsilon_{22} - \lambda) \left( 1 - \norm{\varphi_2}^2 \right)
        \\
        \int F \varphi_2 - \frac 12 \varepsilon_{12} \left( 1 - \norm{\varphi_2}^2 \right)
        \\
        \int G \varphi_1 - \frac 12 \varepsilon_{12} \left( 1 - \norm{\varphi_1}^2 \right)
        \\
        - \int \varphi_1 \varphi_2
    \end{pmatrix}
\end{multline}
We search for orthonormal orbitals $\varphi_1, \varphi_2$.
Therefore, the determinant of this system
should stay
close to $- \varepsilon_{11} - \varepsilon_{22}$ during each Newton's update,
provided $\lambda = 0$.
In particular, the matrix should be invertible for each Newton's iteration $w = w_n$.
The integrals in the right hand side of \eqref{energy_update_system_shifted}
do not depend on $\lambda$
and they are computed by Formulas
\eqref{integral_F_phi_1}-\eqref{integral_G_phi_1}.

It is left to rewrite the $L^2$-equations as
\begin{multline}
\label{shifted_kinetic_delta_phi_1}
    \left(
        - \frac {c_1^2}2 \Delta
        -
        \varepsilon_{11}
        +
        \lambda
    \right)
    \delta \varphi_1
    +
    \left(
        1
        -
        \frac {\delta \theta \sin 2\theta}{c_1^2}
    \right)
    \left(
        - \frac {c_1^2}2 \Delta
        - \varepsilon_{11}
        +
        \lambda
    \right)
    \varphi_1
    \\
    +
    c_1^2
    \left(
        V_{\text{nuc}}
        +
        J (11)
    \right)
    \delta \varphi_1
    +
    2 c_1^2
    J (1 \delta 1)
    \varphi_1
    +
    c_1 c_2
    J (\delta 12)
    \varphi_2
    +
    c_1 c_2
    J (1 \delta 2)
    \varphi_2
    +
    c_1 c_2
    J (12)
    \delta \varphi_2
    -
    \varepsilon_{12} \delta \varphi_2
    \\
    -
    \delta \varepsilon_{12} \varphi_2
    -
    \varepsilon_{12} \varphi_2
    -
    \left(
        (
            \varepsilon_{11}
            -
            \lambda
        )
        \frac {\delta \theta \sin 2\theta}{c_1^2}
        +
        \delta \varepsilon_{11}
        +
        \lambda
    \right)
    \varphi_1
    \\
    +
    \left(
        c_1^2
        - \delta \theta \sin 2\theta
    \right)
    \left(
        V_{\text{nuc}}
        +
        J(11)
    \right)
    \varphi_1
    +
    (    
        c_1 c_2
        +
        \delta \theta  
        \cos 2 \theta
    )
    J (12)
    \varphi_2
    =
    0
\end{multline}
and
\begin{multline}
\label{shifted_kinetic_delta_phi_2}
    \left(
        -
        \frac {c_2^2}2 \Delta
        -
        \varepsilon_{22}
        +
        \lambda
    \right)
    \delta \varphi_2
    +
    \left(
        1
        +
        \frac {\delta \theta \sin 2\theta}{c_2^2}
    \right)
    \left(
        - \frac {c_2^2}2 \Delta
        - \varepsilon_{22}
        +
        \lambda
    \right)
    \varphi_2
    \\
    +
    c_1 c_2
    J (\delta 12)
    \varphi_1
    +
    c_1 c_2
    J (12)
    \delta \varphi_1
    +
    2 c_2^2
    J (2 \delta 2)
    \varphi_2
    +
    c_1 c_2
    J (1 \delta 2)
    \varphi_1
    -
    \varepsilon_{12} \delta \varphi_1
    +
    c_2^2
    \left(
        V_{\text{nuc}}
        +
        J(22)
    \right)
    \delta \varphi_2
    \\
    -
    \delta \varepsilon_{12} \varphi_1
    -
    \varepsilon_{12} \varphi_1
    +
    \left(    
        (
            \varepsilon_{22}
            -
            \lambda
        )
        \frac {\delta \theta \sin 2\theta}{c_2^2}
        -
        \delta \varepsilon_{22}
        -
        \lambda
    \right)    
    \varphi_2
    \\
    +
    \left(
        c_2^2
        +
        \delta \theta \sin 2\theta
    \right)
    \left(
        V_{\text{nuc}}
        +
        J(22)
    \right)
    \varphi_2
    +
    (
        c_1 c_2
        +
        \delta \theta  
        \cos 2 \theta
    )
    J (12)
    \varphi_1
    =
    0
    .
\end{multline}

Finally,
introducing
the resolvents at shifted orbital energies
\[
    R_i^{\lambda}
    =
    \left(
        - \frac {c_i^2}2 \Delta - \varepsilon_{ii} + \lambda
    \right) ^{-1}
    =
    \frac 2{c_i^2} H
    \left(
        \sqrt{ \frac {2 (\lambda - \varepsilon_{ii})}{c_i^2} }
    \right)
    =
    \frac 2{c_i^2} H
    \left(
        \mu_i
    \right)
    =
    \frac 2{c_i^2}
    \left(
        - \Delta + \mu_i^2
    \right) ^{-1}
\]
we can rewrite
Equations \eqref{shifted_kinetic_delta_phi_1}, \eqref{shifted_kinetic_delta_phi_2}
as
\begin{equation}
\begin{aligned}
    \delta \varphi_1
    =
    -
    \left(
        1
        -
        \frac {\delta \theta \sin 2\theta}{c_1^2}
    \right)
    \varphi_1
    -
    R_1^{\lambda} \mathfrak F_1^{\lambda}
    ,
    \\
    \delta \varphi_2
    =
    -
    \left(
        1
        +
        \frac {\delta \theta \sin 2\theta}{c_2^2}
    \right)
    \varphi_2
    -
    R_2^{\lambda} \mathfrak F_2^{\lambda}
    ,
\end{aligned}
\end{equation}
where
\begin{multline*}
    \mathfrak F_1^{\lambda}
    =
    -
    \varepsilon_{12} \delta \varphi_2
    -
    (
        \delta \varepsilon_{12}
        +
        \varepsilon_{12}
    )
    \varphi_2
    -
    \left(
        (
            \varepsilon_{11}
            -
            \lambda
        )
        \frac {\delta \theta \sin 2\theta}{c_1^2}
        +
        \delta \varepsilon_{11}
        +
        \lambda
    \right)
    \varphi_1
    \\
    +
    \left(
        V_{\text{nuc}}
        +
        J (11)
    \right)
    \left(
        c_1^2
        \delta \varphi_1
        +
        \left(
            c_1^2
            - \delta \theta \sin 2\theta
        \right)
        \varphi_1
    \right)
    +
    J (12)
    (
        c_1 c_2
        \delta \varphi_2
        +
        (    
            c_1 c_2
            +
            \delta \theta  
            \cos 2 \theta
        )
        \varphi_2
    )
    \\
    +
    2 c_1^2
    J (1 \delta 1)
    \varphi_1
    +
    c_1 c_2
    J (\delta 12)
    \varphi_2
    +
    c_1 c_2
    J (1 \delta 2)
    \varphi_2
\end{multline*}
and
\begin{multline*}
    \mathfrak F_2^{\lambda}
    =
    -
    \varepsilon_{12} \delta \varphi_1
    -
    (
        \delta \varepsilon_{12}
        +
        \varepsilon_{12}
    )
    \varphi_1
    +
    \left(    
        (
            \varepsilon_{22}
            -
            \lambda
        )
        \frac {\delta \theta \sin 2\theta}{c_2^2}
        -
        \delta \varepsilon_{22}
        -
        \lambda
    \right)    
    \varphi_2
    \\
    +
    \left(
        V_{\text{nuc}}
        +
        J(22)
    \right)
    \left(
        c_2^2
        \delta \varphi_2
        +
        \left(
            c_2^2
            +
            \delta \theta \sin 2\theta
        \right)
        \varphi_2
    \right)
    +
    J (12)
    (
        c_1 c_2
        \delta \varphi_1
        +
        (
            c_1 c_2
            +
            \delta \theta  
            \cos 2 \theta
        )
        \varphi_1
    )
    \\
    +
    2 c_2^2
    J (2 \delta 2)
    \varphi_2
    +
    c_1 c_2
    J (\delta 12)
    \varphi_1
    +
    c_1 c_2
    J (1 \delta 2)
    \varphi_1
    .
\end{multline*}

For $\lambda = 0$
the final Newton system obviously coincides with
the system derived in the previous section.
The importance of defining the Newton's function consistently,
namely, as $\mathcal R = \nabla \mathcal L$,
is obvious,
provided one uses the Levenberg–Marquardt damping $\lambda \geqslant 0$
or any other special technique tailored for Newton's optimization.
It is worth to point out,
that as long as $\lambda$ is small,
\(
    \left(
        V_{\text{nuc}}
        +
        J(ii)
        -
        \lambda
    \right)
    \varphi_i
    \approx
    \left(
        V_{\text{nuc}}
        +
        J(ii)
    \right)
    \varphi_i
    .
\)
Therefore,
the damping affects directly only the diagonal orbital energies $\varepsilon_{ii}$.
In fact,
one can use a different $\lambda_i$ for every energy $\varepsilon_{ii}$.
This justifies the trick of controlling and correcting the sign of $\varepsilon_{ii}$
during numerical simulations,
from the abstract optimization perspective.
Keeping this in mind below, we omit the use of the damping in the formulas.
In actual simulations we modify energies diagonal orbital energies $\varepsilon_{ii}$,
as long as we encounter a positive value or if Newton's step fails providing a lower energy value.

\section{General case}
\setcounter{equation}{0}

The wave function is represented as a sum of $M + 1$ closed shell determinants
\begin{equation}
\label{general_natural_expansion}
    \Psi = \sum_{m = 0}^M c_m \left| m \overline{m} \right \rangle
\end{equation}
with orthonormal spatial orbitals $\varphi_m$ and opposite spins.
This is a natural expansion in the sense that it does not lead to loss of generality,
as was first explained by Löwdin and Shull \cite{Lowdin_Shull1956}.
In other words, a wave function $\Psi$ of rank $M + 1$ for a closed shell molecule
can always be rewritten in the form \eqref{general_natural_expansion}.
In Appendix \ref{Optimality_natural_expansion_section}
we provide a full proof of this claim from the first principles, namely,
from the closed-shell assumption
\(
    \hat{S}^2 \Psi = 0
    .
\)
It is worth to point out that
the representation \eqref{general_natural_expansion} was derived by Löwdin and Shull,
but not from the wave function zero spin condition.
They assumed that the wave function already has a representation,
where each Slater determinant has two spin orbitals with opposite spins.
Then they deduced the orthogonality of the corresponding spatial orbitals.
Therefore, we complement their derivation significantly.

For the natural expansion \eqref{general_natural_expansion}
it is easy to check that the total energy takes the quadratic form
\[
    E
    =
    \left \langle
        \Psi
        \left| \widehat H \right|
        \Psi
    \right \rangle
    =
    \sum_{k, m = 0}^M
    H_{km} c_k c_m
    ,
\]
where the energy matrix elements are
\[
    H_{km}
    =
    \left \langle
        k \overline{k}
        \left| \widehat H \right|
        m \overline{m}
    \right \rangle
    =
    2 \delta_{km} (k|h|m)
    +
    (km|km)
    .
\]
We introduce the Lagrangian
\[
    \mathcal L
    =
    \frac 14 E
    -
    \frac 14 \varepsilon \left( \sum_{m = 0}^M c_m^2 - 1 \right)
    -
    \frac 12 \sum_{i, j  = 0}^M \varepsilon_{ij} \left( \int \varphi_i \varphi_j - \delta_{ij} \right)
\]
with symmetric matrix coefficients $\varepsilon_{ij}$.
It is a function of the variables
\(
    \varphi_0, \ldots, \varphi_M,
    \varepsilon,
    c_0, \ldots, c_M
\)
and $\varepsilon_{ij}$ with $i \leqslant j$.
We compute its gradient $\nabla \mathcal L$ and organize its components as follows.
First, the variational derivatives and then the partial derivatives:
\begin{align}
    \label{nabla_L_1}
    \frac{\delta \mathcal L}{\delta \varphi_k}
    &
    =
    c_k^2
    h \varphi_k
    +
    c_k
    \sum_{m = 0}^M
    c_m
    J(km)
    \varphi_m
    -
    \sum_{m = 0}^M
    \varepsilon_{km} \varphi_m
    , \quad
    k = 0, \ldots, M
    ,
    \\
    \frac{\partial \mathcal L}{\partial \varepsilon}
    &
    =
    -
    \frac 14 \left( \sum_{m = 0}^M c_m^2 - 1 \right)
    ,
    \\
    \frac{\partial \mathcal L}{\partial c_k}
    &
    =
    \frac 12
    \sum_{m = 0}^M
    H_{km} c_m
    -
    \frac 12
    \varepsilon c_k
    , \quad
    k = 0, \ldots, M
    ,
    \\
    \frac{\partial \mathcal L}{\partial \varepsilon_{kk}}
    &=
    - \frac 12 \left( \norm{\varphi_k}^2 - 1 \right)
    , \quad
    k = 0, \ldots, M
    ,
    \\
    \label{nabla_L_5}
    \frac{\partial \mathcal L}{\partial \varepsilon_{ij}}
    &=
    - \int \varphi_i \varphi_j
    , \quad
    0 \leqslant i < j \leqslant M
    .
\end{align}

Now let us calculate the Hessian $d \nabla \mathcal L$.
We differentiate \eqref{nabla_L_1} with respect to orbitals as
\begin{equation}
    \sum_{j = 0}^M
    \partial_{\varphi_j}
    \frac{\delta \mathcal L}{\delta \varphi_k}
    \delta \varphi_j
    =
    c_k^2
    h \delta \varphi_k
    -
    \sum_{m = 0}^M
    \varepsilon_{km} \delta \varphi_m
    +
    c_k
    \sum_{m = 0}^M
    c_m
    \left(
        J(\delta km)
        \varphi_m
        +
        J(k \delta m)
        \varphi_m
        +
        J(km)
        \delta \varphi_m
    \right)
    ,
\end{equation}
where
\(
    k = 0, \ldots, M
    .
\)
Next we continue differentiating \eqref{nabla_L_1} as
\[
    \frac{\partial}{\partial \varepsilon}
    \frac{\delta \mathcal L}{\delta \varphi_k}
    =
    0
    ,
\]
\[
    \sum_{j = 0}^M
    \frac{\partial}{\partial c_j}
    \frac{\delta \mathcal L}{\delta \varphi_k}
    \delta c_j
    =
    2 c_k \delta c_k
    h \varphi_k
    +
    \delta c_k
    \sum_{m = 0}^M
    c_m J(km) \varphi_m
    +
    c_k
    \sum_{m = 0}^M
    \delta c_m J(km) \varphi_m
    ,
\]
\[
    \sum_{0 \leqslant i \leqslant j \leqslant M}
    \frac{\partial}{\partial \varepsilon_{ij}}
    \frac{\delta \mathcal L}{\delta \varphi_k}
    \delta \varepsilon_{ij}
    =
    -
    \sum_{m = 0}^M
    \delta \varepsilon_{km} \varphi_m
    ,
\]
where we extend the orbital energy update by symmetry.
Summing these equalities,
we obtain
\begin{multline}
\label{Newton_LHS1}
\tag{LHS1}
    d
    \frac{\delta \mathcal L}{\delta \varphi_k}
    \delta w
    =
    c_k^2
    h \delta \varphi_k
    -
    \sum_{m = 0}^M
    \varepsilon_{km} \delta \varphi_m
    +
    c_k
    \sum_{m = 0}^M
    c_m
    \left(
        J(\delta km)
        \varphi_m
        +
        J(k \delta m)
        \varphi_m
        +
        J(km)
        \delta \varphi_m
    \right)
    \\
    +
    2 c_k \delta c_k
    h \varphi_k
    +
    \delta c_k
    \sum_{m = 0}^M
    c_m J(km) \varphi_m
    +
    c_k
    \sum_{m = 0}^M
    \delta c_m J(km) \varphi_m
    -
    \sum_{m = 0}^M
    \delta \varepsilon_{km} \varphi_m
    ,
\end{multline}
which are the first $M + 1$ lines of the left had side of the Newton system.
Next,
we have
\begin{equation}
\tag{LHS2}
    d
    \frac{\partial \mathcal L}{\partial \varepsilon}
    \delta w
    =
    -
    \frac 12 \sum_{m = 0}^M c_m \delta c_m
    .
\end{equation}
We compute
\begin{equation*}
    \sum_{j = 0}^M
    \partial_{\varphi_j}
    \frac{\partial \mathcal L}{\partial c_k}
    \delta \varphi_j
    =
    2 c_k (\delta k | h | k )
    +
    \sum_{m = 0}^M
    c_m ( \delta k m | k m )
    +
    \sum_{m = 0}^M
    c_m ( k \delta m | k m )
    ,
\end{equation*}
\begin{equation*}
    \frac{\partial }{\partial {\varepsilon}}
    \frac{\partial \mathcal L}{\partial c_k}
    \delta \varepsilon
    =
    -
    \frac 12 \delta \varepsilon c_k
\end{equation*}
and
\begin{equation*}
    \sum_{j = 0}^M
    \frac{\partial }{\partial c_j}
    \frac{\partial \mathcal L}{\partial c_k}
    \delta c_j
    =
    \frac 12
    \sum_{m = 0}^M
    H_{km} \delta c_m
    -
    \frac 12
    \varepsilon \delta c_k
    .
\end{equation*}
Summing these equalities, we obtain
\begin{equation}
\tag{LHS3}
    d
    \frac{\partial \mathcal L}{\partial c_k}
    \delta w
    =
    2 c_k (\delta k | h | k )
    +
    \sum_{m = 0}^M
    c_m ( \delta k m | k m )
    +
    \sum_{m = 0}^M
    c_m ( k \delta m | k m )
    -
    \frac 12 \delta \varepsilon c_k
    +
    \frac 12
    \sum_{m = 0}^M
    H_{km} \delta c_m
    -
    \frac 12
    \varepsilon \delta c_k
    .
\end{equation}
Finally, the last parts of the Hessian have the form
\begin{equation}
\tag{LHS4}
    d
    \frac{\partial \mathcal L}{\partial \varepsilon_{kk}}
    \delta w
    =
    - \int \varphi_k \delta \varphi_k
    , \quad
    k = 0, \ldots, M
    ,
\end{equation}
\begin{equation}
\label{Newton_LHS5}
\tag{LHS5}
    d
    \frac{\partial \mathcal L}{\partial \varepsilon_{ij}}
    \delta w
    =
    -
    \int \delta \varphi_i \varphi_j
    -
    \int \varphi_i \delta \varphi_j
    , \quad
    0 \leqslant i < j \leqslant M
    .
\end{equation}
Expressions \eqref{Newton_LHS1}-\eqref{Newton_LHS5}
define the Hessian and form the left-hand side of the Newton equation.
The right-hand side is $- \nabla \mathcal L$
with the gradient components given by \eqref{nabla_L_1}-\eqref{nabla_L_5}.
This completes the detailed description of the Newton equation \eqref{general_newton}
with $\mathcal R = \nabla \mathcal L$.
We now rewrite it in the self-consistent form \eqref{general_self_consistent_form},
which is suitable for an iterative procedure.

\subsection{Self-consistent form}

We rewrite the Newton equation in the form
\begin{equation}
\label{general_self_consistent_newton}
    \delta \varphi_k
    =
    \mathbb F_k( \delta \varphi_0, \ldots, \delta \varphi_M; w )
    , \quad
    k = 0, \ldots, M
    ,
\end{equation}
where
\(
    w = ( \varphi_1, \ldots, \varepsilon_{M-1,M} )
\)
is held fixed at each Newton step,
as follows.
We can isolate a subsystem of the form
\[
    \begin{pmatrix}
        0 & c_0 & \cdots & c_M
        \\
        c_0 & & &
        \\
        \vdots  & & \varepsilon - H &
        \\
        c_M & & & 
    \end{pmatrix}
    \begin{pmatrix}
        \delta \varepsilon
        \\
        \delta c_0
        \\
        \vdots
        \\
        \delta c_M
    \end{pmatrix}
    =
    \begin{pmatrix}
        - \frac 12 \left( \sum_{m = 0}^M c_m^2 - 1 \right)
        \\
        f
    \end{pmatrix}
\]
in Newton's equation.
Shortly,
\[
    \begin{pmatrix}
        0 & \mathbf{c}^T
        \\
        \mathbf{c} & \varepsilon - H
    \end{pmatrix}
    \begin{pmatrix}
        \delta \varepsilon
        \\
        \delta \mathbf c
    \end{pmatrix}
    =
    \begin{pmatrix}
        - \frac 12 \left( \sum_{m = 0}^M c_m^2 - 1 \right)
        \\
        f
    \end{pmatrix}
    ,
\]
where
\(\mathbf{c} = (c_0, c_1, \dots, c_M)^T\) is a column vector
and
\(\varepsilon - H\) is an \((M+1) \times (M+1)\) matrix.
On the right-hand side we have a column vector $f$ with the components
\[
    f_k
    =
    \sum_{m = 0}^M
    H_{km} c_m
    -
    \varepsilon c_k
    +
    4 c_k (\delta k | h | k )
    +
    2 \sum_{m = 0}^M
    c_m ( \delta k m | k m )
    +
    2 \sum_{m = 0}^M
    c_m ( k \delta m | k m )
    , \quad
    k = 0, \ldots, M
    .
\]

Further on, we introduce the matrix
\begin{multline*}
    F_{kj}
    =
    \int
    \left(
        d
        \frac{\delta \mathcal L}{\delta \varphi_k}
        \delta w
        +
        \frac{\delta \mathcal L}{\delta \varphi_k}
        +
        \sum_{m = 0}^M
        \delta \varepsilon_{km} \varphi_m
        +
        \sum_{m = 0}^M
        \varepsilon_{km} \delta \varphi_m
    \right)
    \varphi_j
    =
    c_k^2 ( \delta k |h| j )
    \\
    +
    c_k
    \sum_{m = 0}^M
    c_m
    \left(
        ( \delta k m | j m )
        +
        ( k \delta m | j m )
        +
        ( j \delta m | k m )
    \right)
    +
    2 c_k \delta c_k ( k |h| j )
    +
    \sum_{m = 0}^M
    (
        c_k \delta c_m + \delta c_k c_m
    )
    ( j m | k m )
    \\
    +
    c_k^2 ( k |h| j )
    +
    c_k
    \sum_{m = 0}^M
    c_m
    ( j m | k m )
    -
    \underbrace{
        \sum_{m = 0}^M
        \varepsilon_{km} \int \varphi_j \varphi_m
    }_{
        = \varepsilon_{jk}
    }
    .
\end{multline*}
So far the derivation of the self-consistent form was generic.
In order to simplify the equations we will assume from now on,
that the orbitals are normalized as
\(
    \int \varphi_j \varphi_m = \delta_{jm}
    .
\)
Note that this normalization is needed to accelerate the Newton optimization,
as we have seen above.
This also simplifies the equations for the orbital energy updates $\delta \varepsilon_{kj}$
into the following matrix form
\[
    X + \mathcal E Y = F
    ,
\]
where $\mathcal E = ( \varepsilon_{kj} )$,
$X = ( \delta \varepsilon_{kj} )$ is symmetric
and $Y = \left( \int \delta \varphi_j \varphi_k \right)$ is antisymmetric.
As long as the spectra of $\mathcal E$ and $- \mathcal E$ are disjoint,
there exists a unique solution $X$.
For this it is enough to have all the eigenvalues of $\mathcal E$ being negative.
One naturally anticipates the orbital energies to be negative. 
Taking the transpose of both sides
and accounting for $X^T = X$, $Y^T = -Y$
one obtains
\[
    X - Y \mathcal E^T = F^T
    .
\]
Now, we add and subtract the original equation in order to simplify to two separate equations:
$$ 2X = (F + F^T) - \mathcal E Y + Y \mathcal E^T, $$
$$ \mathcal E Y + Y \mathcal E^T = (F - F^T). $$
The equation for $Y$ is a Sylvester equation,
which can be solved by the Bartels–Stewart algorithm
implemented in many software packages.
Its computational cost is $\mathcal O \left( M^3 \right)$ arithmetical operations,
which can be viewed as negligible. 
Once $Y$ has been found, we solve for $X$:
$$ X = \frac{1}{2} (F + F^T + \mathcal E Y^T + Y \mathcal E^T). $$
This ensures that $X = ( \delta \varepsilon_{kj} )$ remains symmetric.
The second matrix $Y$ is not needed by itself.

It remains to precondition the first $M + 1$ equations in the Newton system
\[
    d
    \frac{\delta \mathcal L}{\delta \varphi_k}
    \delta w
    +
    \frac{\delta \mathcal L}{\delta \varphi_k}
    =
    0
\]
by first isolating the term
\(
    - \frac { c_k^2 }2 \Delta - \varepsilon_{kk}
\)
in
\begin{multline*}
    \left(
        - \frac { c_k^2 }2 \Delta - \varepsilon_{kk}
    \right)
    \delta \varphi_k
    +
    \left(
        1
        +
        \frac{ 2 \delta c_k }{c_k}
    \right)
    \left(
        - \frac { c_k^2 }2 \Delta - \varepsilon_{kk}
    \right)
    \varphi_k
    +
    \left(
        \frac{ 2 \delta c_k }{c_k}
        \varepsilon_{kk}
        -
        \delta \varepsilon_{kk} 
    \right)
    \varphi_k
    \\
    +
    c_k^2
    V_{\text{nuc}}
    \left(
        \left(
            1
            +
            \frac{ 2 \delta c_k }{c_k}
        \right)
        \varphi_k
        +
        \delta \varphi_k
    \right)
    -
    \sum_{\substack{m=0 \\ m \ne k}}^M
    \left(
        \varepsilon_{km} \delta \varphi_m
        +
        \delta \varepsilon_{km} \varphi_m
        +
        \varepsilon_{km} \varphi_m
    \right)
    \\
    +
    \delta c_k
    \sum_{m = 0}^M
    c_m J(km) \varphi_m
    +
    c_k
    \sum_{m = 0}^M
    \delta c_m J(km) \varphi_m
    +
    c_k
    \sum_{m = 0}^M
    c_m
    J(km)
    (
        \varphi_m + \delta \varphi_m
    )
    \\
    +
    c_k
    \sum_{m = 0}^M
    c_m
    \left(
        J(\delta km)
        \varphi_m
        +
        J(k \delta m)
        \varphi_m
    \right)
    =
    0
\end{multline*}
and then applying the resolvent $R_k$.
This leads to
\begin{equation}
\label{self_consistent_update}
    \delta \varphi_k
    =
    -
    \left(
        1
        +
        \frac{ 2 \delta c_k }{c_k}
    \right)
    \varphi_k
    -
    R_k \mathfrak F_k
    ,
\end{equation}
where
\begin{multline*}
    \mathfrak F_k
    =
    \left(
        \frac{ 2 \delta c_k }{c_k}
        \varepsilon_{kk}
        -
        \delta \varepsilon_{kk} 
    \right)
    \varphi_k
    +
    c_k^2
    V_{\text{nuc}}
    \left(
        \left(
            1
            +
            \frac{ 2 \delta c_k }{c_k}
        \right)
        \varphi_k
        +
        \delta \varphi_k
    \right)
    \\
    -
    \sum_{\substack{m=0 \\ m \ne k}}^M
    \left(
        \varepsilon_{km} \delta \varphi_m
        +
        \delta \varepsilon_{km} \varphi_m
        +
        \varepsilon_{km} \varphi_m
    \right)
    +
    \sum_{m = 0}^M
    J(km)
    \left(
        (
            \delta c_k
            c_m
            +
            c_k
            \delta c_m
            +
            c_k
            c_m
        )
        \varphi_m
        +
        c_k
        c_m
        \delta \varphi_m
    \right)
    \\
    +
    c_k
    \sum_{m = 0}^M
    c_m
    \left(
        J(\delta km)
        \varphi_m
        +
        J(k \delta m)
        \varphi_m
    \right)
\end{multline*}
and the convolution operator $R_k$ is defined in \eqref{resolvent}.
This finishes the description of $\mathbb F_k$ in \eqref{general_self_consistent_newton}.

\subsection{Numerics}

\begin{figure}[ht!]
    \centering
    {
	\includegraphics[width=0.7\textwidth]
        {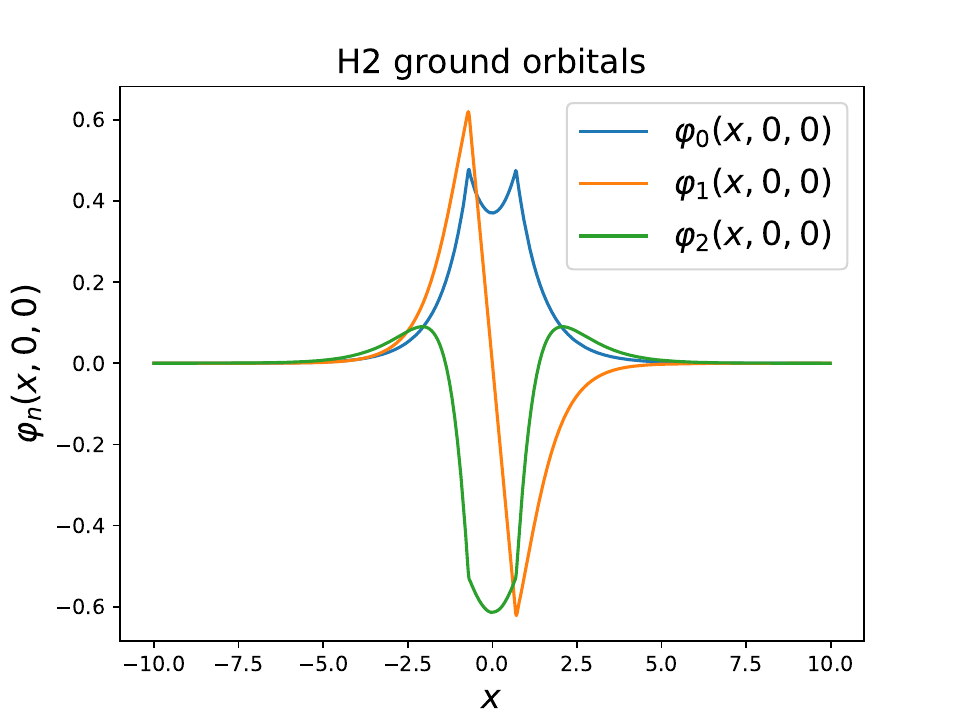}
    }
    \caption
    {
    	Three-determinant ground state of the hydrogen molecule
        with the coefficients
        $c_0 = 0.99253, c_1 = -0.10718$ and $ c_2 = -0.05829$, accordingly.
        The corresponding energy is $-1.15949$.
    }
\label{ground_h2_orbitals_figure}
\end{figure}

The default calculation settings are as follows:
a maximum of 15 iterations for the inner loop
associated with the solution of each Newton equation.
We use \ac{DIIS} with maximum 3 previous iterations.
The initial trust radius is set to 1.0
and modified by dividing by two every time the new iteration results
in a higher energy value.
In addition, we modify the diagonal values of the orbital energies by multiplying by 10.
We initialize the orbital energy matrix by $- \mathds{1}$ at the beginning.
One may reset the orbital energy matrix back to $-\mathds{1}$
at the end of the first few iterations,
in order to avoid unwanted saddle points.

After each iteration the state is normalized using L\"owdin transform.
In addition, CI coefficients are optimized after the orthonormalization,
see Figure \ref{Newton_optimisation_figure}.
These two sub-steps are very cheap compared to solving the Newton system.
The converged result for the two orbital \ac{MCSCF} for the helium atom
with the multiwavelet threshold $\varepsilon_{\text{MRA}} = 10^{-5}$ is
$c_0 = 0.99793, c_1 = -0.06430$
and orbitals symmetric with respect to origin,
shown in Figure \ref{ground_helium_orbitals_figure}.
The corresponding total energy is $-2.87799$.
It is worth comparing with \cite{Nakashima_Nakatsuji},
where the authors report a ground state energy for Helium of -2.90372438136211 a.u. using their free iterative complement interaction (ICI) method.
This result agrees with other high-precision calculations \cite{Aznabaev_Bekbaev_Korobov}.
It is worth noting that, while this is a theoretical calculation,
it is considered extremely accurate
and is often used as a benchmark for experimental measurements and other theoretical methods.

For H$_2$ molecule, we refer to \cite{Kolos_Wolniewicz}:
the equilibrium internuclear distance $R_{\text{nuc}} = 1.4010784$,
the nuclear repulsion corrected total energy $E_{\text{el}} + 1/R_{\text{nuc}}$ is -1.17447.
These values are used as reference values in the present work.
In \cite{Sims_Hagstrom2006} the total energy of
$-1.174475931399$ hartree
at equilibrium $R_{\text{nuc}} = 1.4011$ is reported.
Our three-configuration result yields the energy $-1.15949$
with the coefficients $c_0 = 0.99253, c_1 = -0.10718$ and $ c_2 = -0.05829$.
The corresponding \ac{MCSCF} orbitals are shown in Figure \ref{ground_h2_orbitals_figure}.

\begin{figure}[ht!]
    \centering
    {
	\includegraphics[width=0.7\textwidth]
        {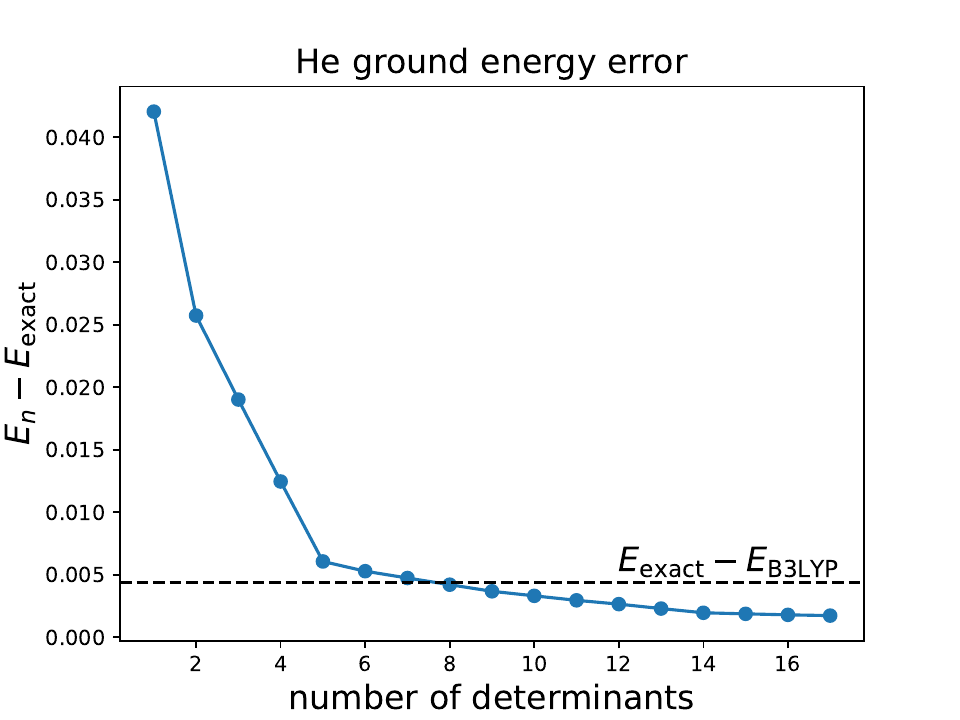}
    }
    \caption
    {
	Convergence of \ac{MCSCF} to the exact ground state energy
        $E_{\text{exact}} = -2.90372$.
    }
\label{ground_helium_error_figure}
\end{figure}

\begin{figure}[ht!]
    \centering
    {
	\includegraphics[width=0.7\textwidth]
        {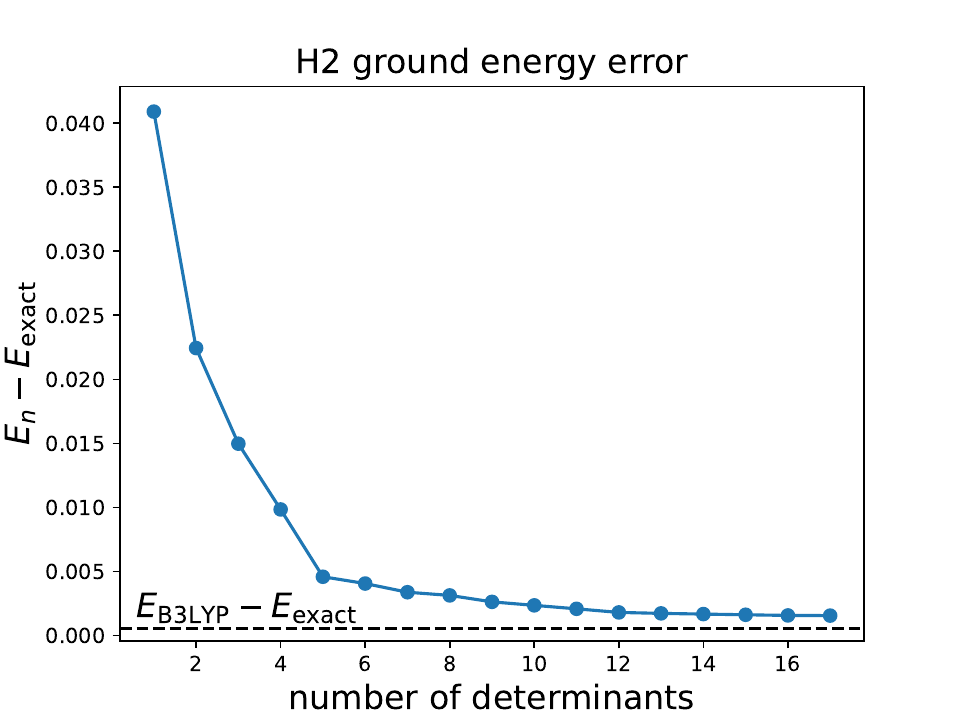}
    }
    \caption
    {
	Convergence of \ac{MCSCF} to the exact ground state energy
        $E_{\text{exact}} = -1.17447$.
    }
\label{ground_h2_error_figure}
\end{figure}

The \ac{MCSCF} approximation approaches the corresponding exact wave function
very slowly as the number of determinants increases,
see Figures \ref{ground_helium_error_figure}, \ref{ground_h2_error_figure}.
For comparison, we also include the calculations with B3LYP functional.
The DFT method provides a lower value than $E_{\text{exact}}$ for He,
and a higher value than $E_{\text{exact}}$ for H$_2$.

It is worth noting that we did not rely on any chemical intuition
in order to choose proper initial guesses for the two quantum systems we regarded.
Instead, we first ran single electron calculations starting with orbitals
combining randomly localized Gaussian functions.
The converged ionic simulations were later used to initialize
the Newton treatment of \ac{MCSCF}.
We follow the same initial guess strategy for the excited states considered below.
Alternatively, one can use spherical harmonics appearing in hydrogen-like atoms.
However, it is not clear how this initialization approach can be extended to larger molecules.
Therefore, we adopt randomized initial calculations.

\subsection{Single-determinant problem}

For the Hartree--Fock problem with $M = 0$,
if the orbital $\varphi_0$ is normalized after each Newton step, then
\[
    \int \varphi_0 \delta \varphi_0 = 0,
\]
which simplifies the scalar subsystem
\[
    \delta \varepsilon
    =
    ( \delta 0 |h| 0 )
    +
    3 ( 0 \delta 0 | 00 )
    +
    ( 0 |h| 0 )
    +
    ( 00 | 00 ) - \varepsilon
    .
\]
Moreover, one can reset the Lagrange multiplier before running Newton's inner loop as
\[
    \varepsilon
    =
    ( 0 |h| 0 )
    +
    ( 00 | 00 ),
\]
in agreement with the Hartree--Fock equation.
Under these assumptions, the Newton system reduces to
\begin{equation*}
    \begin{aligned}
        \delta \varepsilon
        &=
        ( \delta 0 |h| 0 )
        +
        3 ( 0 \delta 0 | 00 ),
        \\
        \delta \varphi
        &=
        -
        \varphi
        +
        \left(
            -\frac{\Delta}2 - \varepsilon
        \right)^{-1}
        \Big[
            \delta \varepsilon \varphi
            -
            \left(
                V_{\text{nuc}}
                +
                J (00)
            \right)
            \left(
                \delta \varphi
                +
                \varphi
            \right)
            -
            2 J (0 \delta 0) \varphi
        \Big]
        .
    \end{aligned}
\end{equation*}
Furthermore,
if one approximates the solution of the Newton system by a single inner iteration,
\[
    \delta \varphi = 0 \;\;\mapsto\;\; \delta \varepsilon = 0 \;\;\mapsto\;\; \delta \varphi,
\]
then the update reduces to
\begin{equation*}
    \varphi
    +
    \delta \varphi
    =
    -
    \left(
        -\frac{\Delta}2 - \varepsilon
    \right)^{-1}
    \Big[
        \left(
            V_{\text{nuc}}
            +
            J (00)
        \right)
        \varphi
    \Big]
    ,
\end{equation*}
which coincides with a well-established Green's function iteration scheme
\cite{Frediani_Fossgaard_Fla_Ruud, Harrison_Fann_Yanai_Gan_Beylkin2004}.
This observation highlights the fundamental role of the resolvent operator and provides further insight into the efficiency of multiwavelet-based methods in electronic structure calculations.

\section{Excited states}
\setcounter{equation}{0}

The first excited singlet state can be approximated as a sum of $M + 1$ closed shell determinants
\[
    \Psi
    =
    \sum_{m = 0}^M c_m \left| m \overline{m} \right \rangle
    .
\]
With the additional constraint, corresponding to the orthogonality of $\Psi$
to the ground state $\Psi_{\text{ground}}$,
the new Lagrangian takes the form
\[
    \mathcal L
    =
    \frac 14 E
    -
    \frac 14 \varepsilon \left( \sum_{m = 0}^M c_m^2 - 1 \right)
    -
    \frac 12 \sum_{i, j  = 0}^M \varepsilon_{ij} \left( \int \varphi_i \varphi_j - \delta_{ij} \right)
    -
    \frac 12 \lambda
    \sum_{m, n  = 0}^{M, M^{\text{gr}}} c_m c_n^{\text{gr}} \left( \int \varphi_m \varphi_n^{\text{gr}} \right) ^2
\]
with symmetric matrix coefficients $\varepsilon_{ij}$.
It is a function of
\(
    \varphi_0, \ldots, \varphi_M,
    \varepsilon,
    c_0, \ldots, c_M
    ,
\)
$\varepsilon_{ij}$ with $i \leqslant j$ and $\lambda$.
Its gradient $\nabla \mathcal L$ consists of the following components
\begin{align}
    \label{nabla_L_1_excited}
    \frac{\delta \mathcal L}{\delta \varphi_k}
    &
    =
    c_k^2
    h \varphi_k
    +
    c_k
    \sum_{m = 0}^M
    c_m
    J(km)
    \varphi_m
    -
    \sum_{m = 0}^M
    \varepsilon_{km} \varphi_m
    -
    \lambda
    \sum_{n  = 0}^{M^{\text{gr}}} c_k c_n^{\text{gr}} \varphi_n^{\text{gr}} \int \varphi_k \varphi_n^{\text{gr}}
    , \quad
    k = 0, \ldots, M
    ,
    \\
    \frac{\partial \mathcal L}{\partial \varepsilon}
    &
    =
    -
    \frac 14 \left( \sum_{m = 0}^M c_m^2 - 1 \right)
    ,
    \\
    \frac{\partial \mathcal L}{\partial c_k}
    &
    =
    \frac 12
    \sum_{m = 0}^M
    H_{km} c_m
    -
    \frac 12
    \varepsilon c_k
    -
    \frac 12 \lambda
    \sum_{n  = 0}^{M^{\text{gr}}} c_n^{\text{gr}} \left( \int \varphi_k \varphi_n^{\text{gr}} \right) ^2
    , \quad
    k = 0, \ldots, M
    ,
    \\
    \frac{\partial \mathcal L}{\partial \varepsilon_{kk}}
    &=
    - \frac 12 \left( \norm{\varphi_k}^2 - 1 \right)
    , \quad
    k = 0, \ldots, M
    ,
    \\
    \frac{\partial \mathcal L}{\partial \varepsilon_{ij}}
    &=
    - \int \varphi_i \varphi_j
    , \quad
    0 \leqslant i < j \leqslant M
    ,
    \\
    \label{nabla_L_6_excited}
    \frac{\partial \mathcal L}{\partial \lambda}
    &=
    -
    \frac 12
    \sum_{m, n  = 0}^{M, M^{\text{gr}}} c_m c_n^{\text{gr}} \left( \int \varphi_m \varphi_n^{\text{gr}} \right) ^2
    .
\end{align}

Next we compute the Hessian $d \nabla \mathcal L$.
We start by differentiating the variational derivatives.
First, with respect to orbitals
\begin{multline}
    \sum_{j = 0}^M
    \partial_{\varphi_j}
    \frac{\delta \mathcal L}{\delta \varphi_k}
    \delta \varphi_j
    =
    c_k^2
    h \delta \varphi_k
    -
    \sum_{m = 0}^M
    \varepsilon_{km} \delta \varphi_m
    +
    c_k
    \sum_{m = 0}^M
    c_m
    \left(
        J(\delta km)
        \varphi_m
        +
        J(k \delta m)
        \varphi_m
        +
        J(km)
        \delta \varphi_m
    \right)
    \\
    -
    \lambda
    \sum_{n  = 0}^{M^{\text{gr}}} c_k c_n^{\text{gr}} \varphi_n^{\text{gr}} \int \delta \varphi_k \varphi_n^{\text{gr}}
    ,
\end{multline}
where
\(
    k = 0, \ldots, M
    .
\)
Then with respect to Lagrange multiplier $\varepsilon$ as
\[
    \frac{\partial}{\partial \varepsilon}
    \frac{\delta \mathcal L}{\delta \varphi_k}
    =
    0
    .
\]
The derivatives with respect to CI coefficients equal
\[
    \sum_{j = 0}^M
    \frac{\partial}{\partial c_j}
    \frac{\delta \mathcal L}{\delta \varphi_k}
    \delta c_j
    =
    2 c_k \delta c_k
    h \varphi_k
    +
    \delta c_k
    \sum_{m = 0}^M
    c_m J(km) \varphi_m
    +
    c_k
    \sum_{m = 0}^M
    \delta c_m J(km) \varphi_m
    -
    \lambda
    \sum_{n  = 0}^{M^{\text{gr}}} \delta c_k c_n^{\text{gr}} \varphi_n^{\text{gr}} \int \varphi_k \varphi_n^{\text{gr}}
    .
\]
The derivatives with respect to orbital energies are
\[
    \sum_{0 \leqslant i \leqslant j \leqslant M}
    \frac{\partial}{\partial \varepsilon_{ij}}
    \frac{\delta \mathcal L}{\delta \varphi_k}
    \delta \varepsilon_{ij}
    =
    -
    \sum_{m = 0}^M
    \delta \varepsilon_{km} \varphi_m
    ,
\]
where we extended the orbital energy update by symmetry.
Finally, the derivative with respect to Lagrange multiplier $\lambda$
responsible for the orthogonality to the ground state
has the following expression
\[
    \frac{\partial}{\partial \lambda}
    \frac{\delta \mathcal L}{\delta \varphi_k}
    \delta \lambda
    =
    -
    \delta \lambda
    \sum_{n  = 0}^{M^{\text{gr}}} c_k c_n^{\text{gr}} \varphi_n^{\text{gr}} \int \varphi_k \varphi_n^{\text{gr}}
    .
\]
Summing these equalities we obtain
the first component
\begin{multline}
\label{Newton_LHS1_excited}
\tag{LHS1}
    d
    \frac{\delta \mathcal L}{\delta \varphi_k}
    \delta w
    =
    c_k^2
    h \delta \varphi_k
    -
    \sum_{m = 0}^M
    \varepsilon_{km} \delta \varphi_m
    +
    c_k
    \sum_{m = 0}^M
    c_m
    \left(
        J(\delta km)
        \varphi_m
        +
        J(k \delta m)
        \varphi_m
        +
        J(km)
        \delta \varphi_m
    \right)
    \\
    +
    2 c_k \delta c_k
    h \varphi_k
    +
    \delta c_k
    \sum_{m = 0}^M
    c_m J(km) \varphi_m
    +
    c_k
    \sum_{m = 0}^M
    \delta c_m J(km) \varphi_m
    -
    \sum_{m = 0}^M
    \delta \varepsilon_{km} \varphi_m
    \\
    -
    \lambda
    \sum_{n  = 0}^{M^{\text{gr}}} c_k c_n^{\text{gr}} \varphi_n^{\text{gr}} \int \delta \varphi_k \varphi_n^{\text{gr}}
    -
    \lambda
    \sum_{n  = 0}^{M^{\text{gr}}} \delta c_k c_n^{\text{gr}} \varphi_n^{\text{gr}} \int \varphi_k \varphi_n^{\text{gr}}
    -
    \delta \lambda
    \sum_{n  = 0}^{M^{\text{gr}}} c_k c_n^{\text{gr}} \varphi_n^{\text{gr}} \int \varphi_k \varphi_n^{\text{gr}}
\end{multline}
of the Hessian $d \nabla \mathcal L(w)$ at the update $\delta w$.
The second component is given by
\begin{equation}
\tag{LHS2}
    d
    \frac{\partial \mathcal L}{\partial \varepsilon}
    \delta w
    =
    -
    \frac 12 \sum_{m = 0}^M c_m \delta c_m
    .
\end{equation}
In order to obtain the third component,
we compute the variations
\begin{equation*}
    \sum_{j = 0}^M
    \partial_{\varphi_j}
    \frac{\partial \mathcal L}{\partial c_k}
    \delta \varphi_j
    =
    2 c_k (\delta k | h | k )
    +
    \sum_{m = 0}^M
    c_m ( \delta k m | k m )
    +
    \sum_{m = 0}^M
    c_m ( k \delta m | k m )
    -
    \lambda
    \sum_{n  = 0}^{M^{\text{gr}}} c_n^{\text{gr}}
    \left( \int \varphi_k \varphi_n^{\text{gr}} \right)
    \int \delta \varphi_k \varphi_n^{\text{gr}}
\end{equation*}
with respect to orbitals
and the partial derivatives
\begin{equation*}
    \frac{\partial }{\partial {\varepsilon}}
    \frac{\partial \mathcal L}{\partial c_k}
    \delta \varepsilon
    =
    -
    \frac 12 \delta \varepsilon c_k
    ,
\end{equation*}
\begin{equation*}
    \sum_{j = 0}^M
    \frac{\partial }{\partial c_j}
    \frac{\partial \mathcal L}{\partial c_k}
    \delta c_j
    =
    \frac 12
    \sum_{m = 0}^M
    H_{km} \delta c_m
    -
    \frac 12
    \varepsilon \delta c_k
    ,
\end{equation*}
\begin{equation*}
    \frac{\partial }{\partial {\lambda}}
    \frac{\partial \mathcal L}{\partial c_k}
    \delta \lambda
    =
    -
    \frac 12 \delta \lambda
    \sum_{n  = 0}^{M^{\text{gr}}} c_n^{\text{gr}} \left( \int \varphi_k \varphi_n^{\text{gr}} \right) ^2
    .
\end{equation*}
Summing these equalities we obtain
the third component
\begin{multline}
\tag{LHS3}
    d
    \frac{\partial \mathcal L}{\partial c_k}
    \delta w
    =
    2 c_k (\delta k | h | k )
    +
    \sum_{m = 0}^M
    c_m ( \delta k m | k m )
    +
    \sum_{m = 0}^M
    c_m ( k \delta m | k m )
    -
    \frac 12 \delta \varepsilon c_k
    +
    \frac 12
    \sum_{m = 0}^M
    H_{km} \delta c_m
    \\
    -
    \frac 12
    \varepsilon \delta c_k
    -
    \lambda
    \sum_{n  = 0}^{M^{\text{gr}}} c_n^{\text{gr}}
    \left( \int \varphi_k \varphi_n^{\text{gr}} \right)
    \int \delta \varphi_k \varphi_n^{\text{gr}}
    -
    \frac 12 \delta \lambda
    \sum_{n  = 0}^{M^{\text{gr}}} c_n^{\text{gr}} \left( \int \varphi_k \varphi_n^{\text{gr}} \right) ^2
    .
\end{multline}
Finally, we compute the remaining components
\begin{equation}
\tag{LHS4}
    d
    \frac{\partial \mathcal L}{\partial \varepsilon_{kk}}
    \delta w
    =
    - \int \varphi_k \delta \varphi_k
    , \quad
    k = 0, \ldots, M
    ,
\end{equation}
\begin{equation}
\tag{LHS5}
    d
    \frac{\partial \mathcal L}{\partial \varepsilon_{ij}}
    \delta w
    =
    -
    \int \delta \varphi_i \varphi_j
    -
    \int \varphi_i \delta \varphi_j
    , \quad
    0 \leqslant i < j \leqslant M
    ,
\end{equation}
\begin{equation}
\label{Newton_LHS6_excited}
\tag{LHS6}
    d
    \frac{\partial \mathcal L}{\partial \lambda}
    \delta w
    =
    -
    \sum_{m, n  = 0}^{M, M^{\text{gr}}} c_m c_n^{\text{gr}}
    \left( \int \varphi_m \varphi_n^{\text{gr}} \right)
    \int \delta \varphi_m \varphi_n^{\text{gr}}
    -
    \frac 12
    \sum_{m, n  = 0}^{M, M^{\text{gr}}} \delta c_m c_n^{\text{gr}} \left( \int \varphi_m \varphi_n^{\text{gr}} \right) ^2
    .
\end{equation}

Expressions \eqref{Newton_LHS1_excited}-\eqref{Newton_LHS6_excited}
define $d \nabla \mathcal L(w)(\delta w)$ staying at the
left-hand side of the Newton equation.
The right hand side is $- \nabla \mathcal L(w)$
with the gradient components \eqref{nabla_L_1_excited}-\eqref{nabla_L_6_excited}.
This finishes the detailed description of the Newton equation.
Next we rewrite it in the self-consistent form,
which will be suitable for an iterative procedure.

\subsection{Self-consistent form}

We rewrite the Newton equation in the form \eqref{general_self_consistent_newton}
where
\[
    w = ( \varphi_1, \ldots, \varphi_M, \varepsilon, c_0, \ldots, \varepsilon_{M-1,M}, \lambda )
\]
is fixed at each Newton step,
as follows.
First, we separate evaluation of
\(
    \delta \varepsilon
    ,
    \delta \lambda
    ,
    \delta \mathbf{c}
\)
as
\[
    \begin{pmatrix}
        0 & 0 & \mathbf{c}^T
        \\
        0 & 0 & \mathbf{v}^T
        \\
        \mathbf{c} & \mathbf{v} & \varepsilon - H
    \end{pmatrix}
    \begin{pmatrix}
        \delta \varepsilon
        \\
        \delta \lambda
        \\
        \delta \mathbf c
    \end{pmatrix}
    =
    \begin{pmatrix}
        - \frac 12 \left( \sum_{m = 0}^M c_m^2 - 1 \right)
        \\
        -
        \langle \mathbf c, \mathbf v \rangle
        -
        2
        \sum_{m, n  = 0}^{M, M^{\text{gr}}} c_m c_n^{\text{gr}}
        \left( \int \varphi_m \varphi_n^{\text{gr}} \right)
        \int \delta \varphi_m \varphi_n^{\text{gr}}
        \\
        f
    \end{pmatrix}
\]
where
\(\mathbf{c} = (c_0, c_1, \dots, c_M)^T\)
and
\(\mathbf{v} = (v_0, v_1, \dots, v_M)^T\)
are column vectors with
\[
    v_k
    =
    \sum_{n  = 0}^{M^{\text{gr}}} c_n^{\text{gr}} \left( \int \varphi_k \varphi_n^{\text{gr}} \right) ^2
    , \quad
    k = 0, \ldots, M
    ,
\]
\(\varepsilon - H\) is an \((M+1) \times (M+1)\) matrix.
The column vector $f$ has the components
\begin{multline*}
    f_k
    =
    \sum_{m = 0}^M
    H_{km} c_m
    -
    \varepsilon c_k
    -
    \lambda v_k
    +
    4 c_k (\delta k | h | k )
    +
    2 \sum_{m = 0}^M
    c_m ( \delta k m | k m )
    +
    2 \sum_{m = 0}^M
    c_m ( k \delta m | k m )
    \\
    -
    2
    \lambda
    \sum_{n  = 0}^{M^{\text{gr}}} c_n^{\text{gr}}
    \left( \int \varphi_k \varphi_n^{\text{gr}} \right)
    \int \delta \varphi_k \varphi_n^{\text{gr}}
    , \quad
    k = 0, \ldots, M
    .
\end{multline*}

The matrix $F$ and the Sylvester equation are defined exactly as above.
One can compute
\begin{multline*}
    F_{kj}
    =
    c_k^2 ( \delta k |h| j )
    +
    c_k^2 ( k |h| j )
    -
    \varepsilon_{kj}
    +
    2 c_k \delta c_k ( k |h| j )
    +
    \sum_{m = 0}^M
    c_k
    c_m
    ( \delta k m + k \delta m | j m )
    \\
    +
    \sum_{m = 0}^M
    \left(
        (
            c_k c_m
            \delta \varphi_m
            +
            ( c_k \delta c_m + \delta c_k c_m + c_k c_m )
            \varphi_m
        )
        \varphi_j | k m
    \right)
    \\
    -
    \lambda c_k
    \sum_{n  = 0}^{M^{\text{gr}}} c_n^{\text{gr}} \int \delta \varphi_k \varphi_n^{\text{gr}}
    \int \varphi_j \varphi_n^{\text{gr}}
    -
    (
        \lambda \delta c_k
        +
        \delta \lambda c_k
        +
        \lambda c_k
    )
    \sum_{n  = 0}^{M^{\text{gr}}} c_n^{\text{gr}} \int \varphi_k \varphi_n^{\text{gr}}
    \int \varphi_j \varphi_n^{\text{gr}}
\end{multline*}
for $k, j = 0, \ldots, M$.

It is left to precondition the first $M + 1$ equations in the Newton system
\[
    d
    \frac{\delta \mathcal L}{\delta \varphi_k}
    \delta w
    +
    \frac{\delta \mathcal L}{\delta \varphi_k}
    =
    0
\]
by inverting the kinetic energy in the equations
\begin{multline*}
    \left(
        - \frac { c_k^2 }2 \Delta - \varepsilon_{kk}
    \right)
    \delta \varphi_k
    +
    \left(
        1
        +
        \frac{ 2 \delta c_k }{c_k}
    \right)
    \left(
        - \frac { c_k^2 }2 \Delta - \varepsilon_{kk}
    \right)
    \varphi_k
    +
    \left(
        \frac{ 2 \delta c_k }{c_k}
        \varepsilon_{kk}
        -
        \delta \varepsilon_{kk} 
    \right)
    \varphi_k
    \\
    +
    c_k^2
    V_{\text{nuc}}
    \left(
        \left(
            1
            +
            \frac{ 2 \delta c_k }{c_k}
        \right)
        \varphi_k
        +
        \delta \varphi_k
    \right)
    -
    \sum_{\substack{m=0 \\ m \ne k}}^M
    \left(
        \varepsilon_{km} \delta \varphi_m
        +
        \delta \varepsilon_{km} \varphi_m
        +
        \varepsilon_{km} \varphi_m
    \right)
    \\
    +
    \delta c_k
    \sum_{m = 0}^M
    c_m J(km) \varphi_m
    +
    c_k
    \sum_{m = 0}^M
    \delta c_m J(km) \varphi_m
    +
    c_k
    \sum_{m = 0}^M
    c_m
    J(km)
    (
        \varphi_m + \delta \varphi_m
    )
    \\
    +
    c_k
    \sum_{m = 0}^M
    c_m
    \left(
        J(\delta km)
        \varphi_m
        +
        J(k \delta m)
        \varphi_m
    \right)
    \\
    -
    \lambda c_k
    \sum_{n  = 0}^{M^{\text{gr}}} c_n^{\text{gr}} \varphi_n^{\text{gr}} \int \delta \varphi_k \varphi_n^{\text{gr}}
    -
    (
        \lambda \delta c_k
        +
        \delta \lambda c_k
        +
        \lambda c_k
    )
    \sum_{n  = 0}^{M^{\text{gr}}} c_n^{\text{gr}} \varphi_n^{\text{gr}} \int \varphi_k \varphi_n^{\text{gr}}
    =
    0
    .
\end{multline*}
Finally, we arrive to \eqref{self_consistent_update} with the new
\begin{multline*}
    \mathfrak F_k
    =
    \left(
        \frac{ 2 \delta c_k }{c_k}
        \varepsilon_{kk}
        -
        \delta \varepsilon_{kk} 
    \right)
    \varphi_k
    +
    c_k^2
    V_{\text{nuc}}
    \left(
        \left(
            1
            +
            \frac{ 2 \delta c_k }{c_k}
        \right)
        \varphi_k
        +
        \delta \varphi_k
    \right)
    \\
    -
    \sum_{\substack{m=0 \\ m \ne k}}^M
    \left(
        \varepsilon_{km} \delta \varphi_m
        +
        \delta \varepsilon_{km} \varphi_m
        +
        \varepsilon_{km} \varphi_m
    \right)
    +
    \sum_{m = 0}^M
    J(km)
    \left(
        (
            \delta c_k
            c_m
            +
            c_k
            \delta c_m
            +
            c_k
            c_m
        )
        \varphi_m
        +
        c_k
        c_m
        \delta \varphi_m
    \right)
    \\
    +
    c_k
    \sum_{m = 0}^M
    c_m
    \left(
        J(\delta km)
        \varphi_m
        +
        J(k \delta m)
        \varphi_m
    \right)
    \\
    -
    \lambda c_k
    \sum_{n  = 0}^{M^{\text{gr}}} c_n^{\text{gr}} \varphi_n^{\text{gr}} \int \delta \varphi_k \varphi_n^{\text{gr}}
    -
    (
        \lambda \delta c_k
        +
        \delta \lambda c_k
        +
        \lambda c_k
    )
    \sum_{n  = 0}^{M^{\text{gr}}} c_n^{\text{gr}} \varphi_n^{\text{gr}} \int \varphi_k \varphi_n^{\text{gr}}
\end{multline*}
and the convolution operator $R_k$ is defined as above.
This finishes the description of $\mathbb F_k$ in \eqref{general_self_consistent_newton}
for the first excited state problem.

It is left to describe the excited L\"owdin step (see Appendix),
and the CI coefficient minimization:
\begin{equation*}
    PHP c = \varepsilon c
    ,
\end{equation*}
where the projection has the elements:
\begin{equation*}
    P_{km}
    =
    \delta_{km}
    -
    \frac{ v_k v_m }{ \norm{v}^2 }
    .
\end{equation*}

\subsection{Numerics}

\begin{figure}[ht!]
    \centering
    {
	\includegraphics[width=0.7\textwidth]
        {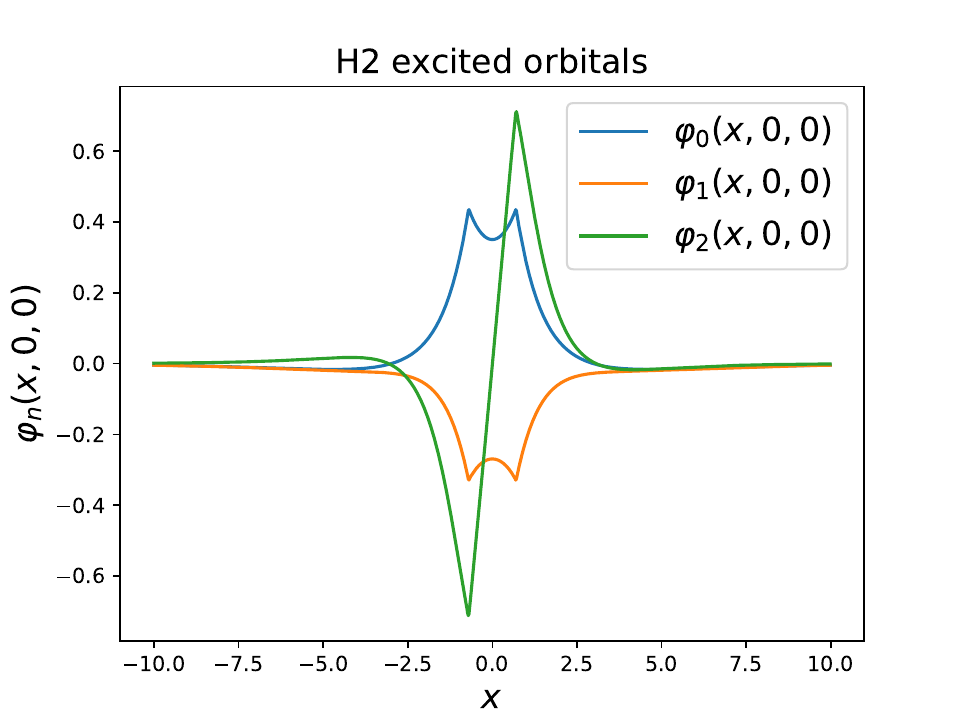}
    }
    \caption
    {
	Three determinants excited state of the hydrogen molecule
        with the coefficients
        $c_0 = 0.76103, c_1 = -0.64796$ and $ c_2 = -0.03128$, accordingly.
        The corresponding energy is $-0.68989$.
    }
\label{excited_h2_orbitals_figure}
\end{figure}

For He atom high accuracy excited states were computed in \cite{Aznabaev_Bekbaev_Korobov}.
The first excited singlet state has the energy
$-2.14597404605441741580502897546192$, that serves as a reference.
Impressively,
our method gives the energy $-2.14350$ for two determinants
associated to coefficients $c_0 = 0.75880, c_1 = -0.65132$
and $M^{\text{gr}} = 5$ corresponding to 6 configurations in the ground state.

For H$_2$ molecule we calculate the first closed shell excited state
at the same internuclear distance $R_{\text{nuc}} = 1.4010784$ as above.
Using 6 configurations in the ground state for the constrain,
we report the energy $-0.68844$ for two configurations with
$c_0 = 0.76135,  c_1 = -0.64834$
and the energy $-0.69028$ for 5 determinants with
$c_0 = 0.76125,  c_1 =  -0.64732,  c_2 =  -0.03140,  c_3 =  0.01669,  c_4 =  0.01461$.
The orbitals can be observed in Figure \ref{excited_h2_orbitals_figure}.
In this case we do not have a high precision energy value, which one could use as a reference,
but some close values were reported in
\cite{Cances_Galicher_Lewin2006, Silkowski_Zientkiewicz_Pachucki2021}.
Figure \ref{excited_h2_orbitals_figure} cannot convey
the orthogonality of the orbitals $\varphi_0$ and $\varphi_1$,
since, to a large extend, their overlap is canceled by contributions from
the long tails.

\appendix
\renewcommand{\theequation}{\thesection.\arabic{equation}}
\section{L\"owdin transformation}
\setcounter{equation}{0}

Given a set of linearly independent orbitals $\phi_1, \ldots, \phi_N \in L^2$
and a non-zero vector $(a_1, \ldots, a_N) \in \mathbb R^N$,
we want to transform them in a way that the new orbitals
$\psi_1, \ldots, \psi_N \in L^2$ and coefficients $b_1, \ldots, b_N \in \mathbb R$
satisfy a particular constrain, while been kept as close as possible to the original ones.
The latter we formalize as the minimization of the distance
\begin{equation}
\label{distance_minimization}
    \sum_{n = 1}^N
    \norm{ \psi_n - \phi_n}^2
    +
    \sum_{n = 1}^N
    \Abs{ b_n - a_n}^2
    .
\end{equation}

\subsection{Ground state constrain}

We impose the following conditions
\[
    \langle \psi_k, \psi_n \rangle
    =
    \delta_{kn}
    , \quad
    \sum_n b_n^2 = 1
\]
on the unknowns $\psi_n, b_n$,
which naturally appear in the ground state MCSCF problem.
And we want to minimize the distance \eqref{distance_minimization}
to the given fixed orbitals $\phi_n$ and coefficients $a_n$.
It turns out that the well-known symmetric L\"owdin orthonormalisation
is the unique solution of this minimization problem \cite{Carlson_Keller}.
Here we re-derive it making use of the Lagrangian formalism.
Firstly, one may easily notice that accounting for the imposed constrain
the objective functional can be simplified.
In other words, the minimization problem is equivalent to the maximization of
\[
    \mathcal F(\psi_1, \ldots, \psi_N, b_1, \ldots, b_N)
    =
    \sum_{n = 1}^N
    \left(
        \langle \phi_n, \psi_n \rangle
        +
        a_n b_n
    \right)
    ,
\]
as pointed out in \cite{Mayer2002}, where one can also find an alternative instructive derivation.

Introducing the Lagrangian
\[
    \mathcal L
    =
    \sum_{n = 1}^N
    \left(
        \langle \phi_n, \psi_n \rangle
        +
        a_n b_n
    \right)
    -
    \frac 12 \sum_{i, j  = 1}^N \varepsilon_{ij}
    \left( \langle \psi_i, \psi_j \rangle - \delta_{ij} \right)
    -
    \frac 12 \beta
    \left(
        \sum_{n = 1}^N b_n^2 - 1
    \right)
\]
with symmetric matrix $\varepsilon = ( \varepsilon_{ij} )$
and computing its gradient,
we arrive to the equations
\[
    \phi_k
    -
    \sum_{n = 1}^N \varepsilon_{kn} \psi_n
    =
    0
\]
\[
    a_k
    -
    \beta b_k
    =
    0
\]
describing stationary points of the functional $\mathcal L(\psi_1, \ldots, \psi_N, b_1, \ldots, b_N)$.
The obtained equations on orbitals are decoupled from the equations on the coefficients,
because they are restricted independently.
The first system implies that the original orbitals $\phi_1, \ldots, \phi_N$ belong to
the span of $\psi_1, \ldots, \psi_N$.
Therefore,
recalling that by the assumption the original orbitals $\phi_1, \ldots, \phi_N$ are linearly independent
we deduce that the matrix $\varepsilon$ is invertible.
The inverse matrix $\varepsilon^{-1}$ is also symmetric, since the corresponding Lagrange multipliers
were introduced symmetrically.
Thus
\[
    \psi_m
    =
    \sum_{k = 1}^N \left( \varepsilon^{-1} \right)_{mk} \phi_k
\]
and from the second system we obtain
\[
    b_k
    =
    \frac 1{\beta}
    a_k
    .
\]
Note that $\beta$ cannot be zero,
otherwise it would violate the non-zero condition imposed on the vector $(a_1, \ldots, a_N)$.
Finally, the Lagrange multipliers can be found from the ground state constrain as
\[
    \delta_{mn}
    =
    \langle \psi_m, \psi_n \rangle
    =
    \sum_{k, l = 1}^N
    \left( \varepsilon^{-1} \right)_{mk}
    \left( \varepsilon^{-1} \right)_{nl}
    \langle \phi_k, \phi_l \rangle
    =
    \left( \varepsilon^{-1} S \varepsilon^{-T} \right)_{mn}
    =
    \left( \varepsilon^{-1} S \varepsilon^{-1} \right)_{mn}
    ,
\]
with $S_{kl} = \langle \phi_k, \phi_l \rangle$,
and
\[
    1
    =
    \norm{b}^2
    =
    \frac 1{\beta^2}
    \norm{a}^2
    .
\]
It leads to $\varepsilon^2 = S$ and $\beta^2 = \norm{a}^2$.
As we will shortly see, $\mathcal F$ achieves a maximum, provided
the signs of $\varepsilon$ and $\beta$ are set as $\varepsilon = \sqrt S$ and $\beta = \norm a$.
The obtained transformation
\begin{equation}
\label{standard_lowdin_orthonormalisation}
    \psi
    =
    S^{-1/2} \phi
    , \quad
    b = a / \norm a
\end{equation}
is the standard L\"owdin orthogonalisation.

It is left to show that the obtained stationary point \eqref{standard_lowdin_orthonormalisation}
is indeed a maximum.
By the Cauchy–Schwarz inequality
\[
    \sum_{n = 1}^N
    a_n b_n
    \leqslant
    \norm a \norm b
    =
    \norm a
\]
for any real vector $b$ with $\norm b = 1$.
Moreover, the equality is achieved if and only if $a, b$ are linearly dependent,
which in this case is equivalent to $b = a / \norm a$.
It shows that the euclidean part of $\mathcal F$
is strictly bounded from above by its value
at the stationary point \eqref{standard_lowdin_orthonormalisation}.
The corresponding functional inner product part of $\mathcal F$
is estimated, 
firstly,
by restricting to the span of the original orbitals $\phi_1, \ldots, \phi_N$.
In other words, we suppose that the new orthonormalised functions $\psi_1, \ldots, \psi_N$
can be obtained by a matrix transformation
\(
    \psi = \alpha \phi
\)
of coordinates $\phi_1, \ldots, \phi_N$.
Obviously, the matrix $\alpha$ is invertible.
From the normalization constrain we deduce
\[
    S = \alpha^{-1} \alpha^{-T}
\]
and so the Lagrange multiplier matrix $\varepsilon$ is related to $\alpha$ as
\[
    \varepsilon
    =
    \sqrt S
    =
    \sqrt{ \alpha^{-1} \alpha^{-T} }
    =
    \Abs{ \alpha^{-1} }
    .
\]
Therefore,
\begin{multline*}
    \sum_{n = 1}^N
    \langle \phi_n, \psi_n \rangle
    =
    \sum_{n, m = 1}^N
    \alpha_{n, m} S_{n, m}
    =
    \trace (\alpha S)
    =
    \trace \left( \alpha^{-1} \right)
    \leqslant
    \norm{ \alpha^{-1} }_1
    =
    \trace \Abs{ \alpha^{-1} }
    \\
    =
    \trace \varepsilon
    =
    \trace (\varepsilon^{-1} S)
    =
    \sum_{n = 1}^N
    \left \langle \phi_n, \left( \varepsilon^{-1} \phi \right)_n \right \rangle
    ,
\end{multline*}
where we have used the well-known relation 
for nuclear operators,
in this particular case for the matrix $\alpha^{-1}$,
between the trace and the norm \cite{Birman_Solomjak}.
It proves the statement in the span of the original orbitals.

Finally, in the general case one can project each orbital $\psi_k$ onto $\Span \phi$ and its
orthogonal complement with the projections $P$ and $Q$, correspondingly,
as
\[
    \psi_k = P \psi_k + Q \psi_k
    ,
\]
where the functions $P \psi_k$ can be obtained from the transformation
\(
    P \psi = \alpha \phi
    .
\)
Thus the orthonormalisation constrain
\[
    \delta_{nm}
    =
    \langle \psi_n, \psi_m \rangle
    =
    \sum_{k,l}
    \alpha_{nk} \alpha_{ml} S_{kl}
    +
    \langle Q \psi_n, Q \psi_m \rangle
\]
leads to
\[
    \mathds{1}
    =
    \alpha S \alpha^T + F
\]
with the overlap matrix $F$
satisfying $0 \leqslant F \leqslant \mathds{1}$.
Indeed, for any vector $x \in \mathbb R^N$ we have
\[
    \langle Fx, x \rangle
    =
    \sum_{n,m} F_{nm} x_n x_m
    =
    \sum_{n,m} \langle Q \psi_n, Q \psi_m \rangle x_n x_m
    =
    \norm{
        Q \left( \sum_n x_n \psi_n \right)
    }^2
    \leqslant
    \norm{
        \sum_n x_n \psi_n
    }^2
    =
    \norm{x}^2
\]
implying the bounds on $F$.
Hence
\[
    \varepsilon^2
    =
    S
    =
    \alpha^{-1} (\mathds{1} - F) \alpha^{-T}
\]
and so
\[
    \varepsilon
    =
    \sqrt S
    =
    \sqrt{ \alpha^{-1} \sqrt{ \mathds{1} - F } \left( \alpha^{-1} \sqrt{ \mathds{1} - F } \right)^T }
    =
    \Abs{ \alpha^{-1} \sqrt{ \mathds{1} - F } }
    .
\]
As above we estimate
\begin{multline*}
    \sum_{n = 1}^N
    \langle \phi_n, \psi_n \rangle
    =
    \sum_{n = 1}^N
    \langle \phi_n, P \psi_n \rangle
    =
    \trace (\alpha S)
    =
    \trace \left( \alpha^{-1} (\mathds{1} - F) \right)
    \leqslant
    \norm{ \alpha^{-1} (\mathds{1} - F) }_1
    \\
    \leqslant
    \norm{ \alpha^{-1} \sqrt{ \mathds{1} - F } }_1
    \norm{ \sqrt{ \mathds{1} - F } }
    \leqslant
    \trace \varepsilon
    =
    \trace (\varepsilon^{-1} S)
    =
    \sum_{n = 1}^N
    \left \langle \phi_n, \left( \varepsilon^{-1} \phi \right)_n \right \rangle
    ,
\end{multline*}
where we used the norm estimate
\(
    \norm{ \sqrt{ \mathds{1} - F } }
    \leqslant
    1
\)
following from
\[
    \norm{
        \sqrt{ \mathds{1} - F } x
    }^2
    =
    \langle (\mathds{1} - F) x, x \rangle
    \leqslant
    \norm{x}^2
\]
holding true for any vector $x \in \mathbb R^N$, obviously.
This completes the proof of the minimization property
of the L\"owdin transform \eqref{standard_lowdin_orthonormalisation}.
For an alternative rigorous exposition we refer to \cite{Carlson_Keller}.

\subsection{Excited state constrain}

We want to maximize
\[
    \sum_{n = 1}^N
    \left(
        \langle \phi_n, \psi_n \rangle
        +
        a_n b_n
    \right)
\]
where $\phi_n, a_n$ with $n = 1, \ldots, N$ are fixed given values.
The unknowns $\psi_n, b_n$ satisfy the following constraints
\[
    \langle \psi_k, \psi_n \rangle
    =
    \delta_{kn}
    , \quad
    \sum_n b_n^2 = 1
\]
and
\[
    \sum_{n = 1}^N \sum_{m = 1}^M b_n c_m
    \langle \psi_n, g_m \rangle ^2
    =
    0
    ,
\]
where $g_m, c_m$ with $m = 1, \ldots, M$ are fixed given and normalized as
\[
    \langle g_l, g_m \rangle
    =
    \delta_{lm}
    , \quad
    \sum_m c_m^2 = 1
    .
\]
We introduce the Lagrangian
\[
    \mathcal L
    =
    \sum_{n = 1}^N
    \left(
        \langle \phi_n, \psi_n \rangle
        +
        a_n b_n
    \right)
    -
    \frac 12 \sum_{i, j  = 1}^N \varepsilon_{ij}
    \left( \langle \psi_i, \psi_j \rangle - \delta_{ij} \right)
    -
    \frac 12 \beta
    \left(
        \sum_{n = 1}^N b_n^2 - 1
    \right)
    -
    \frac 12 \lambda
    \sum_{n = 1}^N \sum_{m = 1}^M b_n c_m
    \langle \psi_n, g_m \rangle ^2
\]
with symmetric matrix $\varepsilon = ( \varepsilon_{ij} )$
and compute its gradient
\[
    \frac{ \delta \mathcal L}{ \delta \psi_k }
    =
    \phi_k
    -
    \sum_{n = 1}^N \varepsilon_{kn} \psi_n
    -
    \lambda b_k
    \sum_{m = 1}^M c_m \langle \psi_k, g_m \rangle g_m
\]
\[
    \frac{ \partial \mathcal L}{ \partial b_k }
    =
    a_k
    -
    \beta b_k
    -
    \frac 12 \lambda
    \sum_{m = 1}^M c_m
    \langle \psi_k, g_m \rangle ^2
\]
Thus the unknowns $\psi_n, b_n$,
the symmetric matrix $\varepsilon = ( \varepsilon_{ij} )$
and scalars $\beta, \lambda$
satisfy the following system
\begin{align*}
    \phi_k
    -
    \sum_{n = 1}^N \varepsilon_{kn} \psi_n
    -
    \lambda b_k
    \sum_{m = 1}^M c_m \langle \psi_k, g_m \rangle g_m
    &=
    0
    \\
    a_k
    -
    \beta b_k
    -
    \frac 12 \lambda
    \sum_{m = 1}^M c_m
    \langle \psi_k, g_m \rangle ^2
    &=
    0
    \\
    \langle \psi_i, \psi_j \rangle
    &=
    \delta_{ij}
    \\
    \sum_{n = 1}^N b_n^2
    &=
    1
    \\
    \sum_{n = 1}^N \sum_{m = 1}^M b_n c_m
    \langle \psi_n, g_m \rangle ^2
    &=
    0
\end{align*}
It is possible to reduce the problem to the finite dimensional space
by searching for $\psi_n$ in the span of $\phi_n, g_m$ as
\[
    \psi_k
    =
    \sum_{n = 1}^N A_{kn} \phi_n
    +
    \sum_{m = 1}^M B_{km} g_m
    .
\]
Therefore we need to find the matrices $A, B, \varepsilon$ and scalars $\beta, \lambda$.
Define
\[
    \Phi_{kl}
    =
    \langle \phi_k, \phi_l \rangle
    , \quad
    k, l = 1, \ldots, N
\]
\[
    G_{ml}
    =
    \langle g_m, \phi_l \rangle
    , \quad
    m = 1, \ldots, M
    , \quad
    l = 1, \ldots, N
\]
These overlap matrices are fixed and given.
Then
\[
    \langle \psi_k, g_m \rangle
    =
    \sum_{n = 1}^N A_{kn} G_{mn}
    +
    B_{km}
\]
and
\[
    \langle \psi_k, \phi_l \rangle
    =
    \sum_{n = 1}^N A_{kn} \Phi_{nl}
    +
    \sum_{m = 1}^M
    B_{km} G_{ml}
    .
\]
We want to maximize
\[
    \sum_{k = 1}^N
    \left(
        \langle \phi_k, \psi_k \rangle
        +
        a_k b_k
    \right)
    =
    \sum_{k = 1}^N
    \left(
        \sum_{n = 1}^N A_{kn} \Phi_{nk}
        +
        \sum_{m = 1}^M B_{km} G_{mk}
        +
        a_k b_k
    \right)
\]
The orthonormality constrain is imposed on the overlap:
\[
    \langle \psi_k, \psi_l \rangle
    =
    \sum_{n, n' = 1}^N A_{kn} A_{ln'} \Phi_{nn'}
    +
    \sum_{m = 1}^M
    B_{km} B_{lm}
    +
    \sum_{n = 1}^N
    \sum_{m = 1}^M
    (
        A_{kn} B_{lm}
        +
        A_{ln} B_{km}
    )
    G_{mn}
    .
\]
Finally, orthogonality to the ground state reads
\begin{equation*}
    \sum_{n = 1}^N \sum_{m = 1}^M b_n c_m
    \left(
        \sum_{n' = 1}^N A_{nn'} G_{mn'}
        +
        B_{nm}
    \right) ^2
    =
    0
\end{equation*}

The optimization was implemented using the \texttt{scipy.optimize.minimize} routine with method \texttt{SLSQP}, subject to nonlinear equality constraints:
orthonormality of $\psi_n$, unit norm of the scalar vector $b_n$, and a bilinear orthogonality condition involving the functions $g_m$. We set the initial guess as
\[
A^{(0)} = I_N, \quad B^{(0)} = 0, \quad \text{and} \quad \| b^{(0)} \| = 1.
\]
To ensure convergence, we increased the maximum number of iterations to 10,000 and reduced the function tolerance to $10^{-12}$.

\section{Optimality of spin-restricted configuration expansion}
\label{Optimality_natural_expansion_section}
\setcounter{equation}{0}

Here we prove the existence of closed shell natural expansion \eqref{general_natural_expansion}
from first principles.
Our proof is valid for both real and complex valued orbitals.
In order to clearly distinguish between spatial and spin orbitals
in the expressions below,
we avoid using Dirac brackets for Slater determinants, as used above,
for example, in the expression \eqref{general_natural_expansion}.
We begin by recalling some basic facts and notations from the electronic structure theory.

\begin{definition} \mbox{} \\
    A two-component $N$-electron wave function $\Psi$
    is said to be \underline{non-relativistically closed shell} if
    \begin{equation}
        \hat{S}^2\Psi = 0,
    \end{equation}
    where
    \begin{align}
        \hat{S}^2 &:= \hat{\mathbf{S}}\cdot\hat{\mathbf{S}}
        ,
        \\
        \hat{\mathbf{S}} &:= \sum_{n=1}^N \hat{\mathbf{s}}_n
        ,
        \\
        \hat{\mathbf{s}}_n &:= 1\otimes\dots\otimes\underbrace{\hat{\mathbf{s}}}_{\mathclap{n\textnormal{-th position}}}\otimes\dots\otimes1
        ,
        \\
        \hat{\mathbf{s}} &:= \frac{1}{2} \left(\begin{array}{c}
            \sigma_x \\
            \sigma_y \\
            \sigma_z
        \end{array}\right)
    \end{align}
    and $\sigma_x$, $\sigma_y$ and $\sigma_z$ are Pauli matrices.
\end{definition}

\begin{remark} \mbox{} \\
    $\hat{S}^2\Psi = 0$ and $\hat{\boldsymbol{S}}\Psi = \boldsymbol{0}$ are equivalent statements because
    \begin{equation*}
        \hat{S}^2\Psi = 0 \quad\Longrightarrow\quad \left\langle\Psi\middle|\hat{S}^2\Psi\right\rangle = 0 \quad\Longrightarrow\quad \left\|\hat{S}_x\Psi\right\|^2 + \left\|\hat{S}_y\Psi\right\|^2 + \left\|\hat{S}_z\Psi\right\|^2 = 0 \quad\Longrightarrow\quad \hat{\mathbf{S}}\Psi = \mathbf{0}.
    \end{equation*}
\end{remark}

\begin{definition} \mbox{} \\
    The \underline{adjoint} of an antilinear operator $ \mathcal V \overset{\hat{T}}{\longrightarrow} \mathcal V$ is the unique antilinear operator $\hat{T}^\dagger$ such that
    \begin{equation}
        \left\langle\phi_1\middle|\hat{T}\phi_2\right\rangle = \left\langle\phi_2\middle|\hat{T}^\dagger\phi_1\right\rangle
    \end{equation}
    for all $\phi_1,\phi_2\in \mathcal V$.
    Antilinearity means that for any scalar $\alpha$ and
    spin-orbital $\phi \in \mathcal V$
    it holds that $\hat{T} (\alpha \phi) = \overline \alpha \hat{T} \phi$.
\end{definition}

\begin{definition} \mbox{} \\
    Electronic ladder operators $\bigwedge^N\mathcal{V} \overset{\hat{a}_\phi}{\longrightarrow} \bigwedge^{N-1}\mathcal{V}$ and $\bigwedge^{N-1}\mathcal{V} \overset{\hat{a}_\phi^\dagger}{\longrightarrow} \bigwedge^N\mathcal{V}$  are defined by
    \begin{align}
        \hat{a}_\phi^\dagger \Psi ={}& \phi\wedge\Psi
        ,
        \\
        \hat{a}_\phi \bigwedge_{n=1}^N \psi_n ={}& \sum_{n=1}^N (-1)^{n-1} \langle\phi|\psi_n\rangle \bigwedge_{n'\neq n} \psi_{n'},
    \end{align}
    where $\mathcal{V}$ denotes the space of spin orbitals.
    These operators satisfy the anticommutation relations
    \begin{align}
        \{\hat{a}_{\phi_1}, \hat{a}_{\phi_2}\} ={}& 0 \\
        \left\{\hat{a}_{\phi_1}^\dagger, \hat{a}_{\phi_2}^\dagger\right\} ={}& 0 \\
        \left\{\hat{a}_{\phi_1}, \hat{a}_{\phi_2}^\dagger\right\} ={}& \langle\phi_1|\phi_2\rangle.
    \end{align}
\end{definition}

\begin{theorem}[Non-relativistic closed shell two-electron canonical representation] \mbox{} \\
    Let $\Psi$ be a real or complex two-component two-electron wave function of rank $L$, and
    \begin{equation*}
        \hat{S}^2\Psi = 0.
    \end{equation*}
    Then there are real scalars $c_1,\dots,c_L$ and orthonormal orbitals $\psi_1,\dots,\psi_L$
    such that
    \begin{equation}
        \Psi = \sum_{l=1}^L c_l \left(\begin{array}{c}
            \psi_l \\
            0
        \end{array}\right)\wedge\left(\begin{array}{c}
            0 \\
            \psi_l
        \end{array}\right).
    \end{equation}
\end{theorem}
\begin{proof} \mbox{} \\
    The idea of the proof is to associate to $\Psi$ an antilinear operator and to exploit its spectral structure to construct the desired orthonormal orbitals $\psi_l$.
    We begin by defining the following antilinear antihermitian operator
    \begin{align*}
        \mathcal{V} &\overset{\hat{T}}{\longrightarrow} \mathcal{V}
        \\
        \psi &\mapsto \hat{a}_\psi\Psi,
    \end{align*}
    where $\mathcal{V}$ denotes the inner product space of spin orbitals.
    For any Hermitian one-electron operator $\hat{O}$
    and for any spin orbital $\phi\in\mathcal{V}$,
    we have
    \begin{equation*}
        \left\{\hat{O}, \hat{T}\right\}\phi = \hat{a}_\phi \hat{O} \Psi.
    \end{equation*}
    Note that $\hat{O}$ can be naturally extended
    to Fock space $\bigoplus_{N=0}^\infty\bigwedge^N\mathcal{V}$.
    This identity can be verified by observing that for all $\phi_1,\phi_2\in\mathcal{V}$
    we have
    \begin{align*}
        \left\langle\phi_1\middle|\left\{\hat{O},\hat{T}\right\}\phi_2\right\rangle ={}& \left\langle\phi_1\middle|\left(\hat{O}\hat{a}_{\phi_2} + \hat{a}_{\hat{O}\phi_2}\right)\Psi\right\rangle \\
        ={}& \left\langle\left(\hat{a}_{\phi_2}^\dagger\hat{O} + \hat{a}_{\hat{O}\phi_2}^\dagger\right)\phi_1\middle|\Psi\right\rangle \\
        ={}& \left\langle\left(\phi_2\wedge\hat{O}\phi_1 + \hat{O}\phi_2\wedge\phi_1\right)\middle|\Psi\right\rangle \\
        ={}& \left\langle\hat{O}\left(\phi_2\wedge\phi_1\right)\middle|\Psi\right\rangle \\
        ={}& \left\langle\hat{O}\hat{a}_{\phi_2}^\dagger\phi_1\middle|\Psi\right\rangle \\
        ={}& \left\langle\phi_1\middle|\hat{a}_{\phi_2}\hat{O}\Psi\right\rangle.
    \end{align*}
    Since $\hat{S}^2\Psi = 0$,
    which means
    $\hat{\boldsymbol{S}}\Psi = \boldsymbol{0}$,
    and so
    \begin{equation*}
        \left\{\hat{\boldsymbol{S}}, \hat{T}\right\} = \boldsymbol{0}.
    \end{equation*}
    Since $\Psi$ has rank $L$,
    it admits a representation
    \begin{equation*}
        \Psi = \sum_{l=1}^L\phi_{l,1}\wedge\phi_{l,2}
    \end{equation*}
    for some spin orbitals $\{\phi_{l,n}\}_{\substack{l=1,\dots,L \\ n=1,2}}$. Define the finite-dimensional vector space
    \begin{equation*}
        V := \Span\{\phi_{l,n}\}_{\substack{l=1,\dots,L \\ n=1,2}}.
    \end{equation*}
    We claim that
    the spin orbitals $\{\phi_{l,n}\}_{\substack{l=1,\dots,L \\ n=1,2}}$ are linearly independent,
    since if they are not, one of the spin orbitals is a linear combination of the others.
    Without loss of generality we can take that spin orbital to be $\phi_{L,2}$.
    Then
    \begin{equation*}
        \phi_{L,2} = \sum_{l=1}^{L-1} \sum_{n=1}^2 \alpha_{l,n}\phi_{l,n} + \beta\phi_{L,1} \qquad\text{for some scalars } \{\alpha_{l,n}\} \text{ and } \beta,
    \end{equation*}
    and
    \begin{align*}
        \Psi ={}& \sum_{l=1}^{L-1} \phi_{l,1}\wedge\phi_{l,2} + \phi_{L,1}\wedge\left(\sum_{l=1}^{L-1} \sum_{n=1}^2 \alpha_{l,n}\phi_{l,n} + \beta\phi_{L,1}\right) \\
        ={}& \sum_{l=1}^{L-1} \phi_{l,1}\wedge\phi_{l,2} + \alpha_{l,2}\phi_{L,1}\wedge\phi_{l,2} - \alpha_{l,1}\phi_{l,1}\wedge\phi_{L,1} \\
        ={}& \sum_{l=1}^{L-1} (\phi_{l,1} + \alpha_{l,2}\phi_{L,1})\wedge(\phi_{l,2} - \alpha_{l,1}\phi_{L,1}).
    \end{align*}
    This contradicts the fact that $\Rank\Psi = L$, proving the linear independence.
    Since
    \begin{equation*}
        \hat{T}\psi = \sum_{l=1}^L \sum_{n=1}^2 (-1)^{n-1} \langle\psi|\phi_{l,n}\rangle \bigwedge_{\substack{n'=1 \\ n'\neq n}}^2\phi_{l,n'},
    \end{equation*}
    $\Image\hat{T}\subseteq V$, and $\hat{T}$ can be restricted to an operator on $V$.
    We now show that the restriction
    \begin{equation*}
        V \xrightarrow{\hat{T}\big|_V} V
    \end{equation*}
    is invertible.
    It suffices to show that $\Ker\hat{T}\big|_V = \{0\}$.
    Let $\psi\in\Ker\hat{T}\big|_V$.
    Then
    \begin{equation*}
        \sum_{l=1}^L \langle\psi|\phi_{l,1}\rangle\phi_{l,2} - \langle\psi|\phi_{l,2}\rangle\phi_{l,1} = 0.
    \end{equation*}
    Since $\{\phi_{l,n}\}_{\substack{l=1,\dots,L \\ n=1,2}}$ is linearly independent,
    $\psi\perp V$ which implies $\psi = 0$.
    Thus $\hat{T}\big|_V$ is invertible.

    Next, we show that $V$ is an invariant subspace for $\hat{S}_z$.
    Let $\psi\in V$.
    Then from invertibility of $\hat{T}\big|_V$, there is a $\phi\in V$ so that $\psi = \hat{T}\phi$.
    The anticommutation relation gives us that $V$ is an invariant subspace, since
    \begin{equation*}
        \hat{S}_z\psi = \hat{S}_z\hat{T}\phi = -\hat{T}\hat{S}_z\phi \in \Image\hat{T} \subseteq V.
    \end{equation*}
    In particular,
    the fact that $\hat{S}_z$ is diagonalizable in $\mathcal V$
    with spectrum $\left\{-\frac{1}{2}, \frac{1}{2}\right\}$,
    translates to the restriction $\hat{S}_z\big|_V$.

    The result now follows from Theorem \ref{diagonalizationofantilinearantihermitianoperator},
    stated and proved below.
    In order to use Theorem \ref{diagonalizationofantilinearantihermitianoperator},
    it remains to show that
    \begin{align*}
        \dim E_{-\frac{1}{2}}\left(\hat{S}_z\big|_V\right) ={}& \dim E_{\frac{1}{2}}\left(\hat{S}_z\big|_V\right)
        ,
        \\
        \dim\left(E_{-\frac{1}{2}}\left(\hat{S}_z\big|_V\right) + E_{\frac{1}{2}}\left(\hat{S}_z\big|_V\right)\right) ={}& 2\left\lfloor\frac{\dim V}{2}\right\rfloor.
    \end{align*}
    To show $\dim E_{-\frac{1}{2}}\left(\hat{S}_z\big|_V\right) = \dim E_{\frac{1}{2}}\left(\hat{S}_z\big|_V\right)$, we note that
    \begin{equation*}
        \hat{T}\left(E_{-\frac{1}{2}}\left(\hat{S}_z\big|_V\right)\right) \subseteq E_{\frac{1}{2}}\left(\hat{S}_z\big|_V\right),
    \end{equation*}
    from the anticommutation relation and use the antilinear isomorphism
    \begin{align*}
        E_{-\frac{1}{2}}\left(\hat{S}_z\big|_V\right) &\longrightarrow E_{\frac{1}{2}}\left(\hat{S}_z\big|_V\right) \\
        \phi &\mapsto \hat{T}\phi.
    \end{align*}
    To show $\dim\left(E_{-\frac{1}{2}}\left(\hat{S}_z\big|_V\right) + E_{\frac{1}{2}}\left(\hat{S}_z\big|_V\right)\right) = 2\left\lfloor\frac{\dim V}{2}\right\rfloor$,
    we use the fact that $\hat{S}_z\big|_V$ is diagonalizable with spectrum $\left\{-\frac{1}{2}, \frac{1}{2}\right\}$, which means $V = E_{-\frac{1}{2}}\left(\hat{S}_z\big|_V\right) \oplus E_{\frac{1}{2}}\left(\hat{S}_z\big|_V\right)$. Furthermore, $V$ is even-dimensional since $\dim V = 2L$.
    By Theorem \ref{diagonalizationofantilinearantihermitianoperator}, there is an orthonormal basis of spin orbitals $\{\psi_{l,n}\}_{\substack{l=1,\dots,L \\ n=1,2}}$ in $V$, and there are real numbers $c_1\dots,c_L$ so that for all $l=1,\dots,L$
    \begin{align*}
        \hat{T}\psi_{l,1} ={}& c_l\psi_{l,2}
        ,
        \\
        \hat{T}\psi_{l,2} ={}& -c_l\psi_{l,1}
        ,
        \\
        \hat{S}_z\psi_{l,1} ={}& \frac{1}{2}\psi_{l,1}
        ,
        \\
        \hat{S}_z\psi_{l,2} ={}& -\frac{1}{2}\psi_{l,2}.
    \end{align*}
    Since the basis is orthonormal, we have that for all $\phi\in V$
    \begin{equation*}
        \phi = \sum_{l=1}^L \left(\left\langle\psi_{l,1}\middle|\phi\right\rangle\psi_{l,1} + \left\langle\psi_{l,2}\middle|\phi\right\rangle\psi_{l,2}\right).
    \end{equation*}
    Using the special property of the basis, we have that for all $\phi\in V$
    \begin{align*}
        \hat{T}\phi ={}& \sum_{l=1}^L \left(\overline{\left\langle\psi_{l,1}\middle|\phi\right\rangle}\hat{T}\psi_{l,1} + \overline{\left\langle\psi_{l,2}\middle|\phi\right\rangle}\hat{T}\psi_{l,2}\right) \\
        ={}& \sum_{l=1}^L c_l \left(\left\langle\phi\middle|\psi_{l,1}\right\rangle\psi_{l,2} - \left\langle\phi\middle|\psi_{l,2}\right\rangle\psi_{l,1}\right) \\
        ={}& \hat{a}_\phi \sum_{l=1}^L c_l\psi_{l,1}\wedge\psi_{l,2}.
    \end{align*}
    We now have two expressions for $\hat{T}\phi$, which gives the equation
    \begin{equation*}
        \hat{a}_\phi\left(\Psi - \sum_{l=1}^L c_l\psi_{l,1}\wedge\psi_{l,2}\right) = 0 \qquad\forall\phi\in V.
    \end{equation*}
    It is straightforward to show that if an $N$-electron wave function $\Phi\in\bigwedge^N V$ 
    satisfies $\hat{a}_\phi\Phi = 0$ for all $\phi\in V$ then $\Phi = 0$.
    Consequently,
    \begin{equation*}
        \Psi = \sum_{l=1}^L c_l\psi_{l,1}\wedge\psi_{l,2},
    \end{equation*}
    and we are done up to the proof of Theorem \ref{diagonalizationofantilinearantihermitianoperator}.
\end{proof}

\begin{proposition}\label{eigenvalueofantihermitianantilinear} \mbox{} \\
    Let $V$ be an inner product space over a field $\mathbb{K}\in\{\mathbb{R}, \mathbb{C}\}$ and $V\overset{\hat{T}}{\longrightarrow}V$ antilinear and
    \begin{equation}
        \hat{T}^\dagger = -\hat{T}.
    \end{equation}
    Then the only eigenvalue $\hat{T}$ can have in $\mathbb{K}$ is $0$.
\end{proposition}
\begin{proof} \mbox{} \\
    Let $\lambda\in\mathbb{K}$ and $v\in V\setminus\{0\}$ be an eigenpair.
    Then
    \begin{align*}
        \hat{T}v ={}& \lambda v \\
        \hat{T}^2v ={}& \overline{\lambda}\hat{T}v \\
        \hat{T}^2v ={}& |\lambda|^2v \\
        \left\langle \hat{T}^2v\middle|v\right\rangle ={}& |\lambda|^2\langle v|v\rangle \\
        \left\langle \hat{T}^\dagger v\middle|\hat{T}v\right\rangle ={}& |\lambda|^2\langle v|v\rangle \\
        -\left\langle \hat{T}v\middle|\hat{T}v\right\rangle ={}& |\lambda|^2\langle v|v\rangle.
    \end{align*}
    Hence $\lambda = 0$, since otherwise a positive quantity would equal a negative one.
\end{proof}

\begin{theorem}\label{diagonalizationofantilinearantihermitianoperator} \mbox{} \\
    Let $V$ be a finite-dimensional inner product space over the field $\mathbb{K}\in\{\mathbb{R},\mathbb{C}\}$, and let $V\overset{\hat{T}}{\longrightarrow}V$ be an antilinear antihermitian operator, that is
    \begin{equation*}
        \hat{T}^\dagger = -\hat{T},
    \end{equation*}
    and let $V\overset{\hat{S}}{\longrightarrow}V$ be a Hermitian $\mathbb{K}$-linear operator with eigenvalues $-s$ and $s$ so that
    \begin{align*}
        \left\{\hat{T},\hat{S}\right\} ={}& 0 \\
        \dim E_{-s}\big(\hat{S}\big) ={}& \dim E_s\big(\hat{S}\big) \\
        \dim\left(E_{-s}\big(\hat{S}\big) + E_s\big(\hat{S}\big)\right) ={}& 2\left\lfloor\frac{\dim V}{2}\right\rfloor.
    \end{align*}
    Then $V$ has an orthonormal basis
    \begin{equation*}
        \{u_1,v_1,u_2,v_2,\dots,u_N,v_N\}
    \end{equation*}
    if it is even-dimensional or
    \begin{equation*}
        \{u_1,v_1,u_2,v_2,\dots,u_N,v_N,w\}
    \end{equation*}
    if it is odd-dimensional and there are $\lambda_1,\dots,\lambda_N\in\mathbb{R}$ so that for all $n=1,\dots,N$
    \begin{align*}
        \hat{T}u_n ={}& \lambda_nv_n \\
        \hat{T}v_n ={}& -\lambda_nu_n \\
        \hat{S}u_n ={}& su_n \\
        \hat{S}v_n ={}& -sv_n,
    \end{align*}
    and, in the case where $V$ is odd-dimensional,
    \begin{equation*}
        Tw = 0.
    \end{equation*}
\end{theorem}
\begin{proof} \mbox{} \\
    Use induction on $\left\lfloor\frac{\dim V}{2}\right\rfloor\in\mathbb{N}_0$. \\
    \\
    \underline{$\left\lfloor\frac{\dim V}{2}\right\rfloor = 0$:} \\
    If $V$ is even-dimensional, $\dim V = 0$, and there is nothing to show. If $V$ is odd-dimensional, $\dim V = 1$. Then there is a non-zero $w\in V$ and since $Tw\in V$, $Tw = \lambda w$ for some $\lambda\in\mathbb{C}$, which means $Tw = 0$ by Proposition~\ref{eigenvalueofantihermitianantilinear}. \\
    \\
    \underline{Holds for $\left\lfloor\frac{\dim V}{2}\right\rfloor = N$ $\quad\Longrightarrow\quad$ holds for $\left\lfloor\frac{\dim V}{2}\right\rfloor = N+1$:} \\
    \underline{If $\hat{T}^2 = 0$:} \\
    We must have that $\hat{T} = 0$ since
    \begin{equation*}
        \|Tv\|^2 = \langle Tv|Tv\rangle = -\left\langle v\middle|T^2v\right\rangle,
    \end{equation*}
    which means the theorem is proven by setting all $\lambda_n = 0$ and picking any orthonormal eigenbasis of $\hat{S}$ since
    \begin{align*}
        \dim E_{-s}\big(\hat{S}\big) ={}& \dim E_s\big(\hat{S}\big) \\
        \dim\left(E_{-s}\big(\hat{S}\big) + E_s\big(\hat{S}\big)\right) ={}& 2(N+1)
    \end{align*}
    which ensures that $N+1$ basis vectors have eigenvalue $s$ and $N+1$ basis vectors have eigenvalue $-s$. \\
    \underline{If $\hat{T}^2 \neq 0$:} \\
    Since
    \begin{equation*}
        \left\{\hat{T},\hat{S}\right\} = 0,
    \end{equation*}
    we have
    \begin{equation*}
        \left[\hat{T}^2,\hat{S}\right] = 0.
    \end{equation*}
     $E_s\big(\hat{S}\big)$ is an invariant subspace for $\hat{T}^2$ since $\hat{T}^2$ commutes with $\hat{S}$. Since $\hat{T}^2$ is a negative-semidefinite linear operator, $\hat{T^2} \neq 0$ and $E_s\big(\hat{S}\big) \neq \varnothing$, there is a $\mu > 0$ and normalized $u_{N+1}\in E_s\big(\hat{S}\big)$ so that
    \begin{equation*}
        \hat{T}^2u_{N+1} = -\mu u_{N+1}.
    \end{equation*}
    Set
    \begin{align*}
        \lambda_{N+1} :={}& \sqrt{\mu} \\
        v_{N+1} :={}& \frac{1}{\lambda_{N+1}}\hat{T}u_{N+1}.
    \end{align*}
    Then
    \begin{align*}
        \hat{T}u_{N+1} ={}& \lambda_{N+1}v_{N+1} \\
        \hat{T}v_{N+1} ={}& -\lambda_{N+1}u_{N+1} \\
        \|u_{N+1}\| ={}& \|v_{N+1}\| = 1 \\
        \langle u_{N+1}|v_{N+1}\rangle ={}& 0.
    \end{align*}
    From the anticommutation relation, we get
    \begin{align*}
        \hat{S}u_{N+1} ={}& su_{N+1} \\
        \hat{S}v_{N+1} ={}& -sv_{N+1}.
    \end{align*}
    The subspace $U := \Span\{u_{N+1},v_{N+1}\}$ is an invariant subspace for $\hat{T}$. The orthogonal complement $U^\perp$ is also an invariant subspace for $\hat{T}$ since for all $u\in U$ and $v\in U^\perp$
    \begin{equation*}
        \left\langle u\middle|\hat{T}v\right\rangle = -\left\langle v\middle|\hat{T}u\right\rangle = 0.
    \end{equation*}
    $U^\perp$ is an invariant subspace for $\hat{S}$ since $U$ is an invariant subspace for $\hat{S}$ and for all $u\in U$ and $v\in U^\perp$
    \begin{equation*}
        \left\langle u\middle|\hat{S}v\right\rangle = \left\langle \hat{S}u\middle|v\right\rangle = 0.
    \end{equation*}
    To use the induction hypothesis, it remains to show that
    \begin{align*}
        \dim E_{-s}\left(\hat{S}\big|_{U^\perp}\right) ={}& \dim E_s\left(\hat{S}\big|_{U^\perp}\right) \\
        \dim\left(E_{-s}\left(\hat{S}\big|_{U^\perp}\right) + E_s\left(\hat{S}\big|_{U^\perp}\right)\right) ={}& 2N.
    \end{align*}
    In the case $s \neq 0$, this follows from
    \begin{align*}
        E_{-s}\big(\hat{S}\big) ={}& E_{-s}\left(\hat{S}\big|_{U^\perp}\right) \oplus \Span\{v_{N+1}\}
        ,
        \\
        E_s\big(\hat{S}\big) ={}& E_s\left(\hat{S}\big|_{U^\perp}\right) \oplus \Span\{u_{N+1}\}.
    \end{align*}
    In the case $s = 0$, it follows from
    \begin{equation*}
        E_0\big(\hat{S}\big) = E_0\left(\hat{S}\big|_{U^\perp}\right) \oplus \Span\{u_{N+1}, v_{N+1}\}.
    \end{equation*}
    By the induction hypothesis, $U^\perp$ has an orthonormal basis $$\{u_1,v_1,u_2,v_2,\dots,u_N,v_N\}$$ if it is even-dimensional or $$\{u_1,v_1,u_2,v_2,\dots,u_N,v_N,w\}$$ if it is odd-dimensional and there are real numbers $\lambda_1,\dots,\lambda_n$ so that for all $n=1,\dots,N$
    \begin{align*}
        Tu_n ={}& \lambda_nv_n \\
        Tv_n ={}& -\lambda_nu_n \\
        \hat{S}u_n ={}& su_n \\
        \hat{S}v_n ={}& -sv_n,
    \end{align*}
    and if $V$ is odd dimensional,
    \begin{equation*}
        Tw = 0.
    \end{equation*}
    This completes the proof.
\end{proof}

\vskip 0.05in
\noindent
{\bf Acknowledgments.}
{
    We acknowledge support from the Research Council of Norway through its Centres of Excellence scheme (Hylleraas centre, 262695), through the FRIPRO grant ReMRChem (324590),  and from NOTUR -- The Norwegian Metacenter for Computational Science through grant of computer time (nn14654k).
}

\bibliographystyle{acm}
\bibliography{bibliography}

\end{document}